\documentclass[aps,prx,twocolumn,superscriptaddress,longbibliography, nofootinbib]{revtex4-1}
\usepackage[T1]{fontenc}
\usepackage[utf8]{inputenc}
\usepackage[english]{babel} 

\usepackage{amsmath}
\usepackage{multirow}
\usepackage{verbatim}
\usepackage{graphicx}
\usepackage{xcolor}
\usepackage{setspace}
\usepackage{braket}
\usepackage{bbold}
\usepackage{color}

\usepackage[colorlinks,bookmarks=false,citecolor=blue,linkcolor=red,urlcolor=blue]{hyperref}

\usepackage{siunitx}
\usepackage{nicefrac}
\usepackage{tipa}
\usepackage{amssymb}
\usepackage{wasysym}
\usepackage{bbold}
\usepackage[abs]{overpic}
\usepackage{cancel}
\graphicspath{{Figures/}}

\makeatletter
 \definecolor{boxback}{HTML}{FFF8B5}
 
 \definecolor{applegreen}{rgb}{0, 0.5, 0.0}

\newcommand{\uibkexp}{\affiliation{Institut f\"ur Experimentalphysik, Universit\"at Innsbruck, Innsbruck, Austria}}
\newcommand{\uibkth}{\affiliation{Institut f\"ur Theoretische Physik, Universit\"at Innsbruck, Innsbruck, Austria}}
\newcommand{\iqoqi}{\affiliation{Institute for Quantum Optics and Quantum Information of the Austrian Academy of Sciences,  Innsbruck, Austria}}
\newcommand{\aqt}{\affiliation{Alpine Quantum Technologies GmbH, Innsbruck, Austria}}
\newcommand{\unipd}{\affiliation{Dipartimento di Fisica e Astronomia, Universit\`a di Padova \& INFN sezione di Padova}}

\definecolor{smoothred}{HTML}{C5232F}
\definecolor{mygreen}{rgb}{0,0.5,0}
\definecolor{myblue}{rgb}{0,0,0.75}
\definecolor{mymagenta}{cmyk}{0,1,0,0.12}

\DeclareMathOperator{\Tr}{Tr}

\newcommand\blfootnote[1]{%
  \begingroup
  \renewcommand\thefootnote{}\footnote{#1}%
  \addtocounter{footnote}{-1}%
  \endgroup
}

\begin{document}
\title{Probing phases of quantum matter with an ion-trap tensor-network quantum eigensolver}

\author{Michael Meth*}\uibkexp
\author{Viacheslav Kuzmin*}\uibkth \iqoqi
\author{Rick van Bijnen}\uibkth \iqoqi
\author{Lukas Postler}\uibkexp
\author{Roman Stricker}\uibkexp
\author{Rainer Blatt}\uibkexp \iqoqi \aqt
\author{Martin Ringbauer}\uibkexp 
\author{Thomas Monz}\uibkexp \aqt
\author{Pietro Silvi}\uibkexp \unipd
\author{Philipp Schindler}\uibkexp
\blfootnote{* These authors contributed equally to this work.}

\date{\today}

\begin{abstract}
\noindent
Tensor-Network (TN) states are efficient parametric representations of ground states of local quantum Hamiltonians extensively used in numerical simulations. Here we encode a TN ansatz state directly into a quantum simulator, which can potentially offer
an exponential advantage over purely numerical simulation. In particular, we demonstrate the optimization of a quantum-encoded TN ansatz state using a variational quantum eigensolver on an ion-trap quantum computer by preparing the ground states of the extended Su-Schrieffer-Heeger model. The generated states are characterized by estimating the topological invariants, verifying their topological order. Our TN encoding as a trapped ion circuit employs only single-site addressing optical pulses -- the native operations naturally available on the platform. We reduce nearest-neighbor crosstalk by selecting different magnetic sublevels, with well-separated transition frequencies, to encode even and odd qubits.
\end{abstract}

\maketitle

\section{Introduction}
\label{sec:introduction}
\noindent
Quantum phases of matter that do not have a classical counterpart, known as {\it exotic} phases, are pivotal in the development of novel quantum materials and devices~\cite{hasanColloquiumTopologicalInsulators2010, qiTopologicalInsulatorsSuperconductors2011}. Recently, topologically ordered phases have drawn much interest as potential fault-tolerant components for quantum technologies~\cite{satzingerRealizingTopologicallyOrdered, semeghiniProbingTopologicalSpin2021, sompet2021realising}. Yet, our complete understanding of such collective quantum phenomena remains incomplete~\cite{wen2019choreographed}. While analytical treatment supports only few (integrable) examples \cite{AKLT,ToricCode,3DTopological}, numerical studies are limited to small systems due to the exponential growth of the state-space in many-body quantum problems. Numerical approximations are thus used at intermediate and large systems. At these sizes, tensor-network (TN) methods provide a successful entanglement-based approach~\cite{Perez-Garcia2006,Schollwock2011,Orus2014}. TN ansatz states can efficiently represent ground states of bulk-gapped Hamiltonians in one and higher dimensions~\cite{PhysRevLett.100.030504, RevModPhys.82.277, Ge_2016}, including topological insulators~\cite{aguadoTopologyQuantumStates2007,moore2010topological}. In one spatial dimension, TNs $-$ and Matrix Product States (MPS) as special case $-$ are efficient when evaluating observables. However, in higher dimensions,  numerical resources for exact evaluations often increase exponentially in the system size. Various efficient numerical approximate techniques are known~\cite{TRG2007,Cornertransfermatrix}, although the precision of these approaches is often not tunable.
\newline\newline\noindent
Quantum simulators~\cite{georgescuQuantumSimulation2014} provide an alternative pathway towards  understanding many-body quantum phenomena. Recent progress in quantum hardware~\cite{national2019manipulating, altmanQuantumSimulatorsArchitectures2021} enabled state-of-the-art experiments that are at the cusp of outperforming classical computers~\cite{ebadi2021quantum, scholl2021quantum, arute2019quantum, Kokail_2019, mazurenko2017cold,Koepsell2019,Sun2021,Vijayan2020,Holten2021,Nichols2019,Brown2019,Monroe2021,semeghiniProbingTopologicalSpin2021}. However, the current Noisy Intermediate-Scale Quantum (NISQ) devices~\cite{Preskill2018quantumcomputingin} are still fairly limited, either by size, controllability or noise. Quantum-classical algorithms, such as Variational Quantum Eigensolvers (VQE)~\cite{Farhi2014, PeruzzoVQE, OMalley2016, McClean2016, Moll2018}, are protocols designed for NISQ devices~\cite{bharti2021noisy} and, in the near-term, have the potential to outperform either purely quantum or classical approaches. Such hybrid methods aim to solve problems where implementing a given numerical task with specialized quantum resources can provide an advantage over the classical hardware. In this context, NISQ processors can be tailored to represent and sample TN states~\cite{Schon2005,banulsSequentiallyGeneratedStates, pichlerUniversalPhotonicQuantum2017, smithCrossingTopologicalPhase2019a, Kuzmin2020, barrattParallelQuantumSimulation2021, weiGenerationPhotonicMatrix2021}, e.g.~by encoding specific classes of variational TNs as programmable quantum circuits. Said circuits thus allow the implementation of an efficient variational ansatz, used for instance in VQE \cite{McClean2016,OMalley2016}, to prepare ground states of local gapped Hamiltonians on a quantum simulator. The optimization of the circuit parameters is performed via classical computation. Meanwhile, observables (including the energy cost function) are sampled directly on the quantum device, potentially providing an advantage over classical approaches. 

\noindent
In the present work, we utilize trapped ion quantum resources, sketched in Fig.~\ref{fig:fig1_circuit_ions}(a), to implement a tensor-network-based VQE (TN-VQE). Its purpose is to prepare a many-body entangled ground state on the ion qubits. The ions are sequentially interacting with a common motional (phonon) mode, which acts as an entanglement carrier~\cite{PZ_CNOT,Moelmer1999, Porras2004} or quantum data bus (QDB), as highlighted in Fig.~\ref{fig:fig1_circuit_ions}(b).
In our approach, the interactions are optimized by a classical algorithm with the ultimate goal to realize a target many-body quantum state with the phonon mode being disentangled from the ions at the end of the circuit~\cite{Kuzmin2020}.
\newline\newline\noindent
As a relevant target problem, we use our TN-VQE to demonstrate topologically ordered ground-state phases in the non-integrable extended Su–Schrieffer–Heeger model (eSSH)~\cite{SSH,ElbenTopoInvariants}, shown in Fig.~\ref{fig:fig2_overview}(a). The target states allow an accurate, efficient representation as MPS since they are ground states of 1D gapped Hamiltonians~\cite{1DgroundstatesareMPS,Schollwock2011,Orus2014}. We experimentally approximate the ground states of various model phases by running the TN-VQE algorithm. Finally, we measure Many-Body Topological Invariants~\citep{ElbenTopoInvariants} (MBTI) in the prepared states to detect phases and identify their topological order~\cite{haegeman2012order,Pollmann2012}, see Fig.~\ref{fig:fig2_overview}(b) and (c).

\section{tensor-network variational quantum eigensolver with trapped ions}

\begin{figure}[tb]
    \centering
    \includegraphics[scale=0.28]{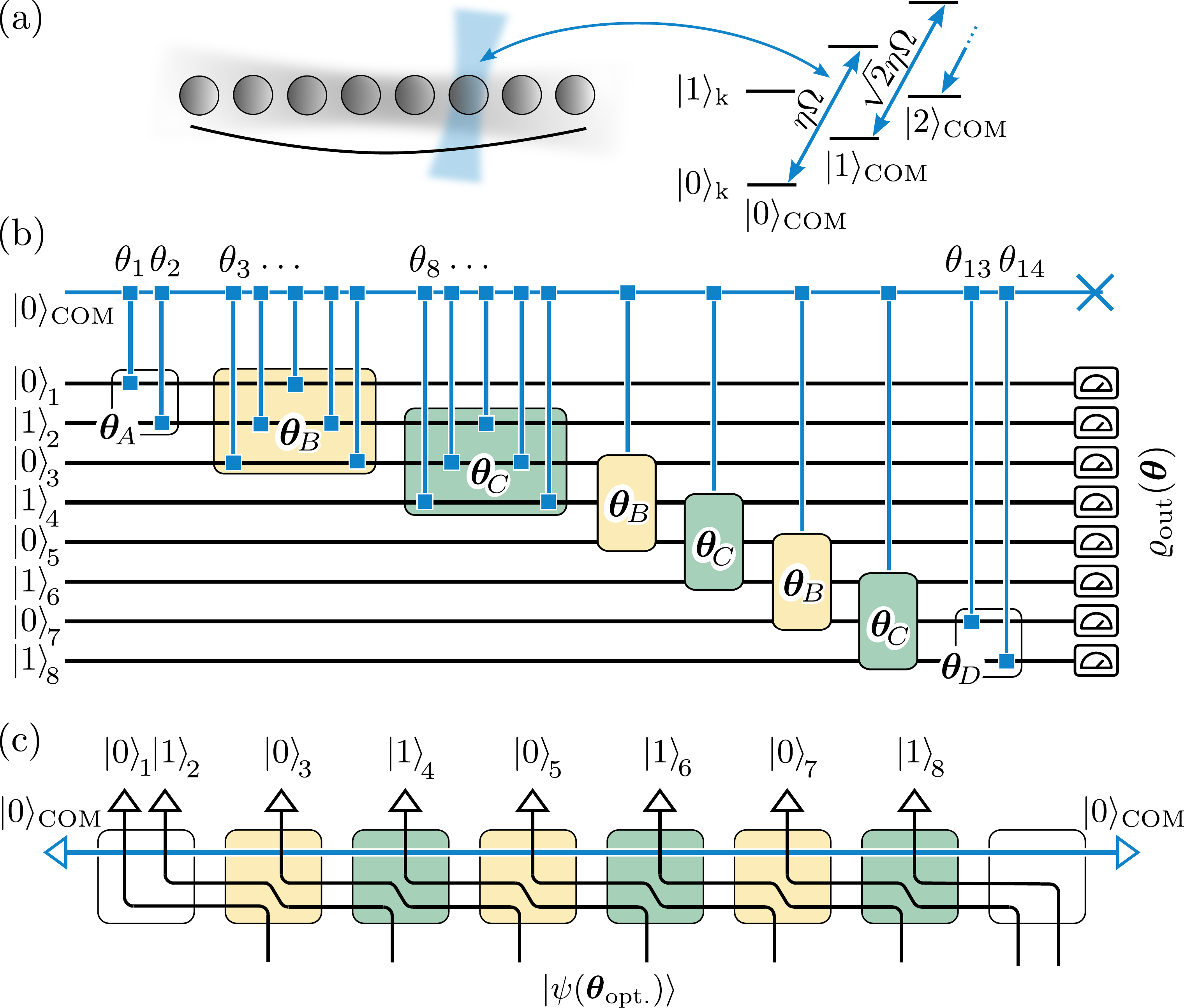}
    \caption{ Scheme of the TN VQE realized on a string of 8 ions. (a)
    The logical states $\ket{0}_k$ and $\ket{1}_k$ are encoded in electronic levels of ion $k$. $\ket{i}_\mathrm{COM}$ are the basis states of the  bosonic axial center-of-mass (COM) mode. We implement (addressed) single qubit operations on the first (blue) motional sideband of the COM mode by laser pulses. (b) Tensor-Network (TN) variational circuit built with blue-sideband operations as the only resource. (c) A pure TN state with the bounded bond dimension generated by the optimized circuit upon disentangling the COM mode from the ions. The empty triangles indicate contraction with pure singe-qubit states written nearby.
    }
    \label{fig:fig1_circuit_ions}
\end{figure}
\noindent
In this section, we briefly review the concept of the TN-VQE approach in trapped ions previously proposed in Ref. \cite{Kuzmin2020}. Further details on our experimental implementation are provided in Sec.~\ref{sec:exper_setup}. We consider a chain of $N$ ions in a linear Paul trap as sketched in Fig.\ref{fig:fig1_circuit_ions}(a). For each site $k$, the qubit $(\ket{0}_k, \ket{1}_k)$ is encoded in a pair of electronic levels of the respective ion. 
The electronic levels can be controllably coupled to collective vibrational (phonon) modes of the ion chain. The entanglement among the ions is distributed by bosonic excitations of the phonon modes, which act as `quantum data buses' (QDBs).
In typical setups, a common approach to create entanglement relies on the simultaneous off-resonant coupling of multiple target ions to the QDBs via a bichromatic field \citep{Moelmer1999, Porras2004}. In this approach, phonons are excited only virtually, which ensures that the QDBs are not entangled with the qubits. However, the underlying process is only at second order in the Lamb-Dike parameter $\eta$, which is typically small $\eta\ll1$ (Lamb-Dicke regime~\citep{gardiner2015quantum}).

Here we consider elementary quantum processing resources that couple qubits and a single phonon mode as a first-order process in $\eta$, thus actively creating entanglement between ions and phonons.
In practice, we employ controlled quantum dynamics generated by the Anti-Jaynes-Cummings Hamiltonian at any single site $k$
\begin{equation}
    H_{k}=i\eta\Omega\left(a\sigma_{k}^{-}-a^{\dagger}\sigma_{k}^{+}\right), 
    \label{eq:anti_JC}
\end{equation}
with Rabi frequency $\Omega$, in the Lamb-Dicke regime. Here $a^\dagger\sigma_k^{+}$ excites the qubit $\ket{0}_k \longmapsto \sigma^{+}\ket{0}_k = \ket{1}_k$ and simultaneously creates a phonon $\ket{n}\longmapsto a^{\dagger} \ket{n}= \sqrt{n+1}\ket{n+1}$, while $a \sigma_k^-$ does the opposite.

\noindent
For optical qubits, such dynamics are realized
by driving with blue-detuned local laser pulses in resonance with the motional sideband of a selected phonon mode~\cite{WinelandSingleTrappedIon2003}
as sketched in Fig.~\ref{fig:fig1_circuit_ions}(a).
Here, we consider the axial center-of-mass (COM) mode due to the near homogeneous coupling to all ions in the chain and the large frequency gap to higher order modes~\citep{James_1998}. Originally, these sideband operations were proposed and used for implementing the C-NOT gate in trapped ions \citep{PZ_CNOT, Schmidt_Kaler_CNOT, Schmidt_Kaler_quantum_gate}. However, the accurate implementation requires precise calibration of the experimental setup as well as cooling of the phonon modes to zero temperature, as the effective Rabi frequency of the sideband operations heavily depends on the phonon population as $\sim\sqrt{n}$. Miscalibrations and any finite phonon population lead to imperfect gates and, as a result, residual entanglement between the QDB and the qubits at the end of a circuit.
Recently, it was proposed~\cite{Kuzmin2020} to tackle this problem by employing adaptive feedback-loop strategies, such as the Variational Quantum Eigensolver (VQE) ~\cite{Farhi2014, PeruzzoVQE, OMalley2016, McClean2016, Moll2018}.

VQE is a NISQ protocol which aims to prepare ground states of interacting Hamiltonians on programmable quantum hardware, while mitigating the imperfections. VQE can address Hamiltonians inaccessible for a given quantum platform and provide the best possible outcome for the available resources. However, identifying the best programmable resources in a specific quantum platform, for a given task, determines the efficiency of VQE. For ion traps, it was shown~\cite{Kuzmin2020} that VQE with sideband operations can suppress the detrimental impact of the finite motional temperatures. The goal is to prepare the pure ground state of a given Hamiltonian $H_\mathrm{targ}$ using the phonon mode as a data bus, which is required to be disentangled (only) at the end of the circuit. To this end, sideband operations are arranged to build a variational circuit $U(\boldsymbol{\theta})=\prod_{l}\mathrm{exp}(-i\theta_{l}H_{k(l)})$, where the $l$-th operation acts on the $k(l)$-th ion, and the parameters $\boldsymbol{\theta}\equiv\{\theta_l\}$ are controlled by the durations of the laser pulses.
For simplicity, let us consider the ideal case where the ions are initialized in some accessible pure state
$\ket{\psi}_\mathrm{in}$, and the QDB is prepared at zero temperature $\ket{0}_\mathrm{COM}$, with
$a |0\rangle_\mathrm{COM} = 0$. Unlike VQE in closed systems, the output variational states of the qubits
\begin{equation}
    \rho_\mathrm{out}(\boldsymbol\theta) = \Tr_\mathrm{COM}
    \left[U(\boldsymbol\theta)
    \left( \ket{\psi} \bra{\psi}_\mathrm{in} \otimes\ket{0} \bra{0}_\mathrm{COM} \right)
    U^{\dagger}(\boldsymbol\theta) \right],
    \label{eq:TN_VQE}
\end{equation}
are not restricted to pure states and are generally mixed due to residual ion-phonon entanglement at the end of the circuit.  Nevertheless, the procedure to optimize the variational parameters is analogous to standard VQE.  Namely, we experimentally measure the energy
\begin{equation}
 E(\boldsymbol\theta) = \Tr[H_\mathrm{targ.}\rho_\mathrm{out}(\boldsymbol\theta)]
\end{equation} 
of the output state for the target model $H_\mathrm{targ.}$, by evaluating all contributing observables.
As usual, several experimental shots are required to evaluate the energy expectation value within a desired errorbar \cite{Kokail_2019}.
The measured energy then acts as a cost function of a numerical variational optimizer, which iteratively proposes new parameter sets $\boldsymbol\theta$ to find a global minimum in the landscape of the variational ansatz manifold. The optimization algorithm we employ is discussed in detail in Sec.~\ref{sec:variational_optimization}. Upon convergence to the ground state energy, for some optimal parameters $\boldsymbol\theta_\mathrm{opt.}$, the prepared state $\rho_\mathrm{out}(\boldsymbol\theta_\mathrm{opt.})$ approaches the manifold of (nearly) degenerate ground states. If the ground state is unique, then $\rho_\mathrm{out}(\boldsymbol\theta)=\ket{\psi(\boldsymbol\theta_\mathrm{opt.})}\bra{\psi(\boldsymbol\theta_\mathrm{opt.})}$, thus qubits are disentangled from the QDB at the end of the preparation. At this optimal scenario, our circuit ansatz guarantees~\cite{Schon2005} that the optimized output state can be represented as
\begin{equation}
\ket{\psi(\boldsymbol\theta_\mathrm{opt.})} = \sum\limits_{j_1\dotsc j_N}\boldsymbol{A}_{j_1}^{[1]}\dotsc\boldsymbol{A}_{j_N}^{[N]}\ket{j_1,\dotsc j_N},
\label{eq:MPS}
\end{equation}
as sketched graphically in Fig.~\ref{fig:fig1_circuit_ions}(c) \cite{Kuzmin2020}.
Here, $j_{k}\in\lbrace 0, 1\rbrace$ are indices of the canonical basis of the $k$-th qubit and $\boldsymbol{A}^{[k]}_0$, $\boldsymbol{A}^{[k]}_1$ are $D_k\times D_{k+1}$ parametric matrices as indicated in Fig.~\ref{fig:fig1_circuit_ions}(c), with $D_1 = D_{N+1} = 1$ under open boundary conditions. States~(\ref{eq:MPS}) are MPS~\cite{MPSfirstpaper,Schon2005}, and the bond dimension $D = \max_k D_k$ is the refinement parameter of the ansatz. The entanglement entropy of the real-space bipartitions in MPS is bounded as $s\leq \log D$, matching the area law of entanglement for the ground states of gapped 1D Hamiltonians~\cite{eisert2010colloquium}. Indeed, it was shown that MPS provides an efficient and faithful ansatz to approximate such states \cite{1DgroundstatesareMPS,Schollwock2011,Orus2014}. In our circuit, $D$ is controlled by the lattice size $m$ of the sideband operation blocks (colored boxes in Fig.~\ref{fig:fig1_circuit_ions}), such that $D\lesssim 2^{m-1}$. 

\noindent
Additionally, typical quantum lattice Hamiltonians $H_\mathrm{targ.}=\sum_k h_k$ exhibit translational invariance, namely $h_k=h_{k+\ell}$ with a period of $\ell$ lattice sites. Ground states of translation-invariant Hamiltonians are well approximated by bulk-translation-invariant MPS with tensors $\boldsymbol{A}^{[k]}_{j_k}=\boldsymbol{A}^{[k+\ell]}_{j_{k+\ell}}$ in the bulk. In our circuit this can be achieved by repeatedly using the \textit{same} subset of variational parameters along the circuit, with period $\ell$: In the language of Fig.~\ref{fig:fig1_circuit_ions}, boxes of the same colors use the same parameters.
The optimal size of the edge blocks with the unique parameters can be identified variationally in the experiment, and it is expected to grow solely with the quantum correlation length of the target state. Consequently, the number of independent scalar parameters in the circuit ansatz does not ultimately grow for increasing system size.
\section{Target model}
\label{Sec:SSH}
\begin{figure*}[tb]
    \centering
    \includegraphics[scale=0.65]{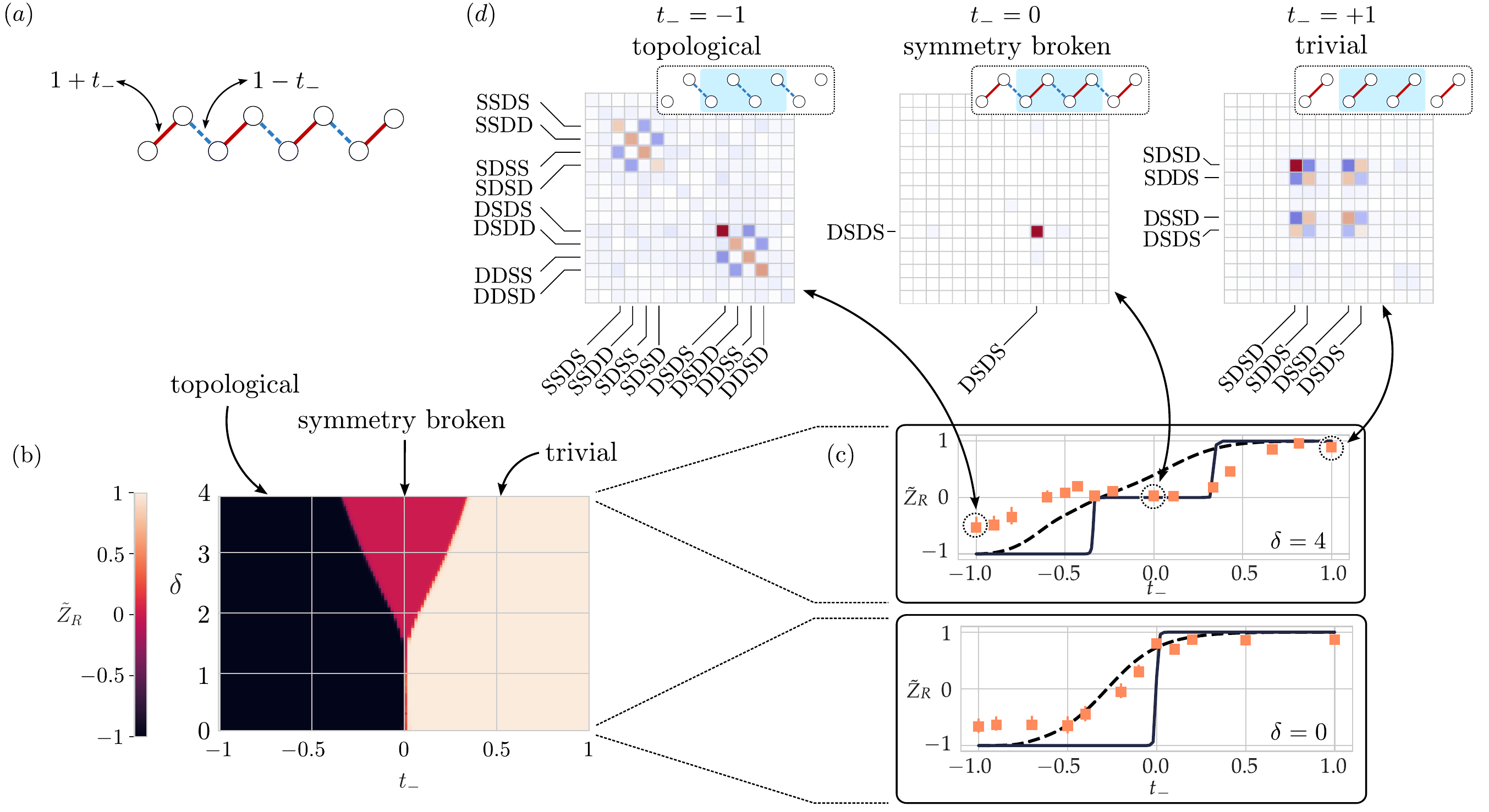}
    \caption{(a) Sketch of the eSSH model for $N=8$ sites, with neighboring sites connected by alternating coupling strengths $1\pm t_{-}$ denoted by solid red and dashed blue bonds. (b)
    Partial-reflection MBTI $\tilde{Z}_R$ [Eq.~(\ref{eq:mbti_zr})] calculated numerically for a system of infinite size $N\to\infty$ and $n=100$ using iDMRG technique~\cite{Schollwock2011}. Labels give the phases of the eSSH model. (c) Comparison of numerical values of MBTI for a finite-sized system (dashed), the thermodynamic limit (solid) and the experimentally obtained data (square markers) for $\delta = 0$ (bottom) and $\delta = 4$ (top). (d) Real component of the reduced density matrices for the four central ions at $\delta = 4$ and $t_{-} \in \{-1, 0, +1\}$. The corresponding data points of $\tilde{Z}_R$ are indicated by arrows.}
    \label{fig:fig2_overview}
\end{figure*}
\noindent
We use TN VQE in an ion trap to study phases in the interacting extension of the Su-Schrieffer-Heeger model \citep{SSH_PRL, PRB_22}, a one-dimensional spin-chain Hamiltonian capturing the transport properties of polymer molecules \citep{Polyacetylene}.
Open boundary conditions are required to exhibit topological order, and the Hamiltonian for $N$ sites reads
\begin{multline}
H_{\text{eSSH}}=\sum_{k=1}^{N-1}\left[1+(-1)^{k-1}t_{-}\right]\\
\times\left(\sigma_{k}^{x}\sigma_{k+1}^{x}+\sigma_{k}^{y}\sigma_{k+1}^{y}+\delta\sigma_{k}^{z}\sigma_{k+1}^{z}\right),
    \label{eq:SSHmodel}
\end{multline}
in dimensionless units, where $\sigma^{\mu}_k$ are the Pauli operators acting on qubit $k$. The coupling $t_{-}$ controls the `staggerization' of the interaction strength, separating even-odd from odd-even pairs as sketched in Fig.~\ref{fig:fig2_overview}(a). Additionally, $\delta$ defines the anisotropy of the XXZ-type interaction, and the `standard' SSH model~\cite{SSH} coincides with the case $\delta = 0$.
For $\delta\neq\{0,1\}$, the model is 
not integrable, thus requiring numerical techniques to study its behavior.
In the antiferromagnetic regime ($\delta \geq 0$), the model exhibits three different energy-gapped phases at zero temperature~\citep{ElbenTopoInvariants}, as indicated in Fig.~\ref{fig:fig2_overview}(b):
{\it (i)} a non-degenerate {\it trivial} dimer phase,
{\it (ii)} a four-fold (quasi-) degenerate {\it Symmetry-Protected Topological} (SPT) dimer phase, with soft excitations localized at the edges, {\it (iii)} a spontaneously {\it symmetry-broken}, Ising antiferromagnetic phase emergent at large $\delta$.

The transition between symmetry protected topological phases can be detected using many-body topological invariants~\cite{haegeman2012order,Pollmann2012} associated with the symmetries protecting the corresponding phases of the 1D system. In case of the eSSH model, these symmetries~\cite{PollmannTopoEntSpectrum,PhysRevB.85.075125} are the dihedral group of $\pi$-rotations about two orthogonal axes, reflection symmetry with respect to the center of the bond, or time-reversal symmetry. In present work, we consider the partial reflection and the partial time-reversal MBTIs~\cite{PollmannsMBTI} corresponding to the two last symmetries. Each of of the MBTIs can be used as a separate phase detector. The partial reflection MBTI reads \citep{PollmannsMBTI, ElbenTopoInvariants}
\begin{equation}
    \tilde{Z}_R = \frac{\mathrm{Tr}[\rho_I \hat{R}_I]}{\sqrt{\left(\mathrm{Tr}[\rho_{I_1}^2] + \mathrm{Tr}[\rho_{I_2}^2]\right)/2}}.
    \label{eq:mbti_zr}
\end{equation}
Here, $\rho_I$, $\rho_{I_1}$, and $\rho_{I_2}$ are reduced states of subsystems $I=I_1\cup I_2$, where $I$ includes an even number $n<N$ of sites in the middle, and $I_1$ and $I_2$ are their left and right subsystems. The parity operator $\hat{R}_I = \hat{R}_I^{\dagger} = \hat{R}_I^{-1}$ reflects the 1D lattice around its center, practically swapping each symmetric pair of qubits. Analogously, the time-reversal MBTI~\cite{PollmannsMBTI, ElbenTopoInvariants} is
\begin{equation}
    \tilde{Z}_{T}=\frac{\mathrm{Tr}[\rho_{I}u_{\text{T}}\rho_{I}^{T_{1}}u_{\text{T}}^{\dagger}]}{\left[\left(\mathrm{Tr}[\rho_{I_{1}}^{2}]+\mathrm{Tr}[\rho_{I_{2}}^{2}]\right)/2\right]^{3/2}},
    \label{eq:mbti_zt}
\end{equation}
where $T_1$ indicates the partial transpose operation on the partition $I_1$, and $u_{T}=\prod_{i\in I_{1}}\sigma_{i}^{y}$, so that the anti-unitary mapping
$\rho_{I} \to u_{\text{T}}\rho_{I}^{T_{1}}u_{\text{T}}^{\dagger}$ completely inverts the spin (thus time-reversal) of each qubit inside $I$.
In the thermodynamic limit, with $N \gg n\to\infty$ (both $N$ and $n$ being multiples of four), $\tilde{Z}_R$ and $\tilde{Z}_T$ approach discrete values in the gapped phases, as demonstrated in Fig.~\ref{fig:fig2_overview}(b) abs (c). Thus they are actual {\it invariants} under any fluctuation which does not disrupt phase properties.
Precisely, they acquire the value
$\tilde{Z}_R =\tilde{Z}_T = 1$ in the trivial phase, $-1$ in the topological phase, and $0$ in the symmetry-broken phase. The subsystem size $n$ required to achieve the convergence depends on the correlation length in the system~\cite{ElbenTopoInvariants}. Thus, in the finite-size system, the measured MBTI become smoothed in proximity of phase boundaries, as shown in Fig.~\ref{fig:fig2_overview}(c) for $\tilde{Z}_R$, where we take $N=8$ and $n=4$.   

\noindent
Far from the phase boundaries, the eSSH ground states are gapped phases, satisfying the entanglement area law~\cite{ RevModPhys.82.277, Ge_2016}, and thus allowing an efficient MPS representation that can be realized with TN-VQE. 
Notably, blue-detuned sideband operations of Eq.~\eqref{eq:anti_JC} are a sufficient resource to prepare ground states of any real, $z$-magnetization conserving Hamiltonian~\cite{Kuzmin2020}. Moreover, these operations take into account specific symmetry properties of $H_{\text{eSSH}}$, eventually simplifying the optimization problem within TN-VQE. Precisely, due to its symmetries, $H_{\text{eSSH}}$ always exhibits at least one ground state with zero $z$-magnetization and all real amplitudes in the canonical basis.
Accordingly, blue-detuned sideband operations realize real-valued variational unitaries $U(\boldsymbol{\theta})$ since $H_k$ are fully imaginary. Furthermore, they protect an `extended' magnetization symmetry $1/2\sum_{k}\sigma_{k}^{z}- a^{\dagger}a$. Thus, if the initial qubit state has well-defined $z$-magnetization, so does the output qubit state once the COM mode has been variationally disentangled \cite{Kuzmin2020}. Thus, by protecting these two symmetries (magnetization, complex conjugation), which are present in the target model, we substantially simplify the variational optimization problem \cite{Kokail_2019}. Conversely, we can not protect the SPTO-symmetries (reflection, time reversal, etc.): doing so would prevent us from establishing topological order from a trivial input state.
\section{Experimental implementation}
\label{sec:exper_setup}
\begin{figure}[b!]
    \centering
    \includegraphics[scale=0.36]{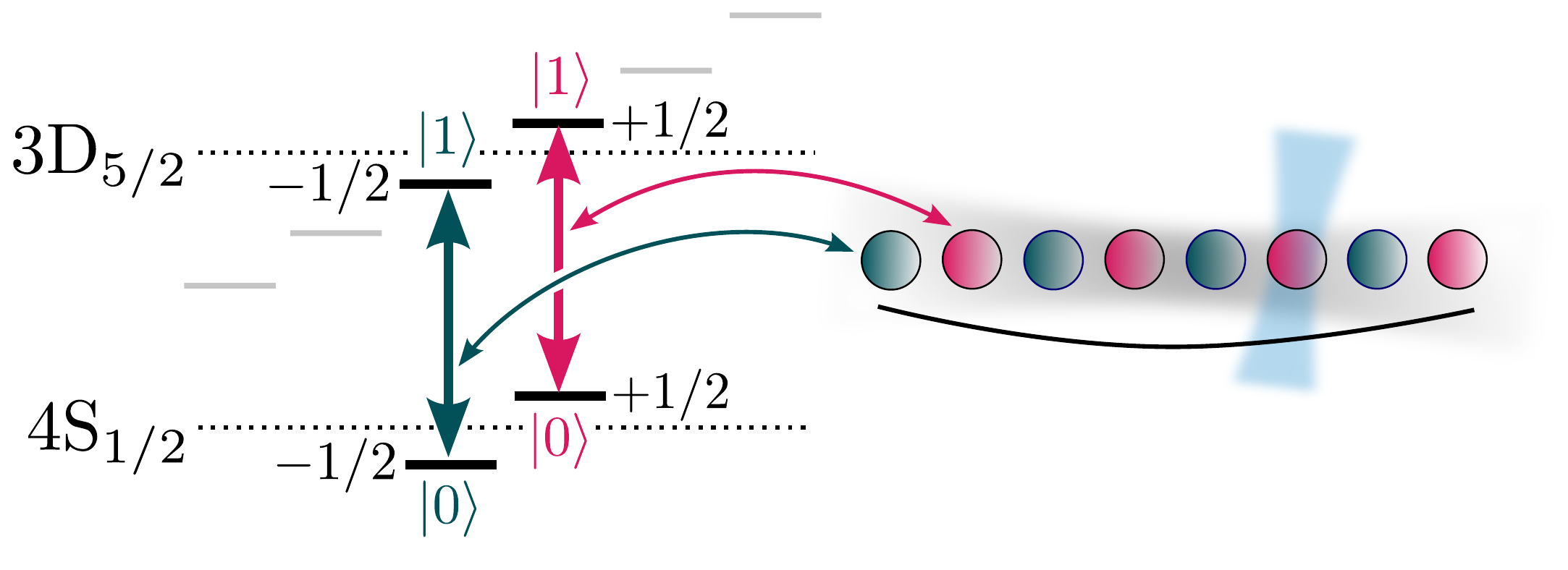}
    \caption{Cross-talk suppression by staggered encoding of qubits in different, well separated Zeeman transitions. For a given site $k$ the qubit is implemented in the $\ket{4S, m = (-1)^k~\nicefrac{1}{2}}\longleftrightarrow\ket{3D, m = (-1)^k~\nicefrac{1}{2}}$ transition.}
    \label{fig:fig3_trapped_ions_scheme}
\end{figure}

\noindent
We implement TN-VQE on an ion trap quantum simulator, using laser-cooled $^{40}\mathrm{Ca}^+$ ions in vacuum \citep{PSchindler2013}. All experimental results presented here are carried out on linear strings of eight ions. 
We encode the qubits in the optical transitions from the electronic ground state $\mathrm{4S}_{\nicefrac{1}{2}}$ to the $\mathrm{3D}_{\nicefrac{5}{2}}$ manifold, such that $\ket{0}=\ket{\mathrm{4S}_{\nicefrac{1}{2}}}$ and $\ket{1}=\ket{\mathrm{3D}_{\nicefrac{5}{2}}}$. The $\mathrm{3D}_{\nicefrac{5}{2}}$ state is long-lived with a life time of $T_1 = \SI{1.15}{s}$.  Initially, the system is prepared in the electronic and motional ground state by a sequence of Doppler-, polarization-gradient, and sideband-cooling and optical pumping \citep{PSchindler2013}; we achieve a mean phonon number $\bar{n} \leq 0.05$.
The quantum state of each qubit is manipulated by a sequence of laser pulses. For each circuit we aim to implement the sideband operations Eq.~(\ref{eq:anti_JC}) on a single qubit at a time, requiring precise control over the position of the laser beam~\citep{2021AQTION_setup, AddressingKim_2008}. In our setup, we address the ions individually via a high numerical aperture objective aligned at an angle of $22.5^\circ$ with respect to the ion string axis. Such single qubit operations will inevitably introduce cross-talk-errors in the non-addressed ions. Due to the geometry of the laser beam, a finite residual field will overlap with -- most prominently -- immediate neighbors, inducing excess rotations and phase errors~\cite{SchindlerDiss}. Comparing the Rabi frequency $\Omega_j$ of a target qubit on site $j$ with its neighbors $j\pm 1$ we measure the ratio $\nicefrac{\Omega_{j\pm 1}}{\Omega_j}$ to be as large as $5.9(5)\%$ in the center of the ion string, which corresponds to a relative laser intensity of $0.35\%$.

\noindent
We seek to minimize the cross-talk errors arising during state preparation and measurement via two schemes. First, resonant errors on the $\ket{S}\longleftrightarrow\ket{D}$ carrier transition are suppressed by implementing a sequence of qubit rotations, decoupling the spectators from the target ion $j$. Here, any rotation $R$ of the qubit by an angle $\theta$ on the Bloch sphere is decomposed into a set of single-qubit rotations
\begin{equation}
    R_j(\theta) = X_j(\nicefrac{\pi}{2}) Z_j(\theta) X_j(\nicefrac{-\pi}{2}),
    \label{eq:U3Gate}
\end{equation}
where $X_j$ rotates the state vector about the x-axis and $Z_j$ about the z-axis. We implement $X_j$ by resonantly driving the carrier transition, while $Z_j$ is realized by a far-detuned laser pulse, inducing an AC Stark shift, in turn rotating the state of the qubit in the equatorial plane of the Bloch sphere. Since the AC Stark shift is proportional to the intensity of the laser rather than the amplitude as in the case of the resonant transition, 
the spectator ions become quadratically decoupled from the operation on the target qubit. Thus, resonant cross-talk can be attributed to the $X$ operations in the sequence in Eq.~(\ref{eq:U3Gate}) alone and as both of them are subjected to the same systematic errors, any resonant error is cancelled out by the opposing phase angles. With this method we manage to suppress resonant cross-talk to well below 1\%.

\noindent
However, this scheme is not applicable when driving sideband operations. First, the Stark shift experienced by the individual states depends on the phonon occupation $n$ \citep{Schmidt_Kaler_2004}. Second, the effective Rabi frequency scales as $\sqrt{n}$, yielding a different evolution for each state. Here we implement a novel scheme using the internal structure of $^{40}\mathrm{Ca}$. We encode the qubit in different, well separated transitions in the $4\mathrm{S}_{\nicefrac{1}{2}}$ and $3\mathrm{D}_{\nicefrac{5}{2}}$ manifolds -- specifically, we select two transitions with magnetic quantum numbers $m = \pm\nicefrac{1}{2}$ as shown in Fig.~\ref{fig:fig3_trapped_ions_scheme}. Thus, for a given site $k$ the encoding is defined as $\ket{0} = \ket{4S, m=(-1)^k~ \nicefrac{1}{2}}$ and $\ket{1} = \ket{3D, m=(-1)^k~\nicefrac{1}{2}}$, respectively. This method prevents any unwanted sideband excitation on neighboring ions. However, it introduces additional efforts during state preparation. For each ion required to be initialized in the $\ket{4S, m = +\nicefrac{1}{2}}$ ground state two additional laser pulses would be needed. Conveniently, as our circuits start out from a N\'eel state $\ket{0101\dotsc}$, we initialize $\ket{3D, m=+\nicefrac{1}{2}}$ on all even sites where only a single pulse on the transition $\ket{4S, m=-\nicefrac{1}{2}}\longleftrightarrow\ket{3D, m=+\nicefrac{1}{2}}$ is required for each respective site. We quantify the benefits of this scheme via state tomography of a circuit with $M = 14$ sideband operations on four qubits using parameters optimized in numerical simulation, given in App. \ref{app:app_four_qubit_sequence}, and compare the data with the state obtained by the circuit simulation. We achieve a fidelity of $\mathcal{F} = 89.7(11)\%$ compared to $82.5(12)\%$ in the case with all qubits encoded in the $\ket{4S, m=-\nicefrac{1}{2}}\longleftrightarrow\ket{3D, m=-\nicefrac{1}{2}}$ transition. Assuming the state fidelity is described by $\mathcal{F} = (\mathcal{F}_{\mathrm{BSB}})^M$, the fidelity of a single sideband operation is given by $\mathcal{F}_{\mathrm{BSB}} = 99.23(7)\%$. 

\noindent
Apart from the state preparation and measurement, the full circuit is implemented using only single-ion sideband operations. In these operations, the phonon-ion coupling scales with the Lamb-Dicke parameter $\eta$, which heavily depends on the trap geometry, or more precisely, on the overlap of the incident laser beam and the trap axis. In our setup we measure $\eta = 0.038$, implying a requirement of up to $2\mathrm{mW}$ of peak laser power to implement the sideband operations, which in turn induce AC stark shifts $\Delta$ by coupling to the carrier transition with a strength on the order of $\Delta\approx\SI{5}{kHz}$. These shifts are actively compensated by a second, far detuned laser beam \cite{Haeffner_2003}. Since this requires substantial power in the compensation beam and with the total available laser power being finite, the achievable Rabi-frequency is also limited. We found the optimum to be at a sideband coupling strength of $\eta\Omega = 2\pi\cdot\SI{8}{kHz}$, such that a $\pi$ rotation $\ket{0, n = 0} \longleftrightarrow \ket{1, n = 1}$ is performed in $\approx\SI{125}{\mu s}$.

In contrast to other entangling schemes like the M\o lmer-S\o rensen gate~\citep{Moelmer1999}, the sideband operations executed in our circuits actively entangle the ions with the phonon mode. Uncontrollable interactions with the environment cause the qubits to depolarize, mainly due to heating and motional dephasing as a consequence of, predominantly, electric field noise in the trap \citep{TrappedIonsWineland}. In our setup, we measure the heating rate of $\Gamma_{\mathrm{H}} = 27(2)$ phonons per second in an 8 ion crystal. Furthermore, we measure the motional coherence time $\tau_{\mathrm{COM}}$ of the COM mode via Ramsey spectroscopy. For a single qubit we obtain $\tau_{\mathrm{COM}} = 101.9(1)\mathrm{ms}$, which is comparable to the coherence time of the laser given by $T_2 = 107(15)\mathrm{ms}$. However, the motional coherence time will decrease with the number of ions in the trap. On an 8 ion string we measure $\tau_{\mathrm{COM}} = 21.9(2)\mathrm{ms}$, which is by a factor of 5 lower than $T_2$. As such, we identify heating and motional dephasing as our main decoherence mechanisms. Consequently, in our setup, it is paramount, that each sideband sequence is finished well within the characteristic times $\tau_{\mathrm{H}} = \nicefrac{1}{\Gamma_{\mathrm{H}}}$ and $\tau_{\mathrm{COM}}$ to ensure the faithful implementation of the target state.
\section{Variational optimization}
\label{sec:variational_optimization}
\begin{figure}[tb]
    \centering
    \includegraphics[scale=0.58]{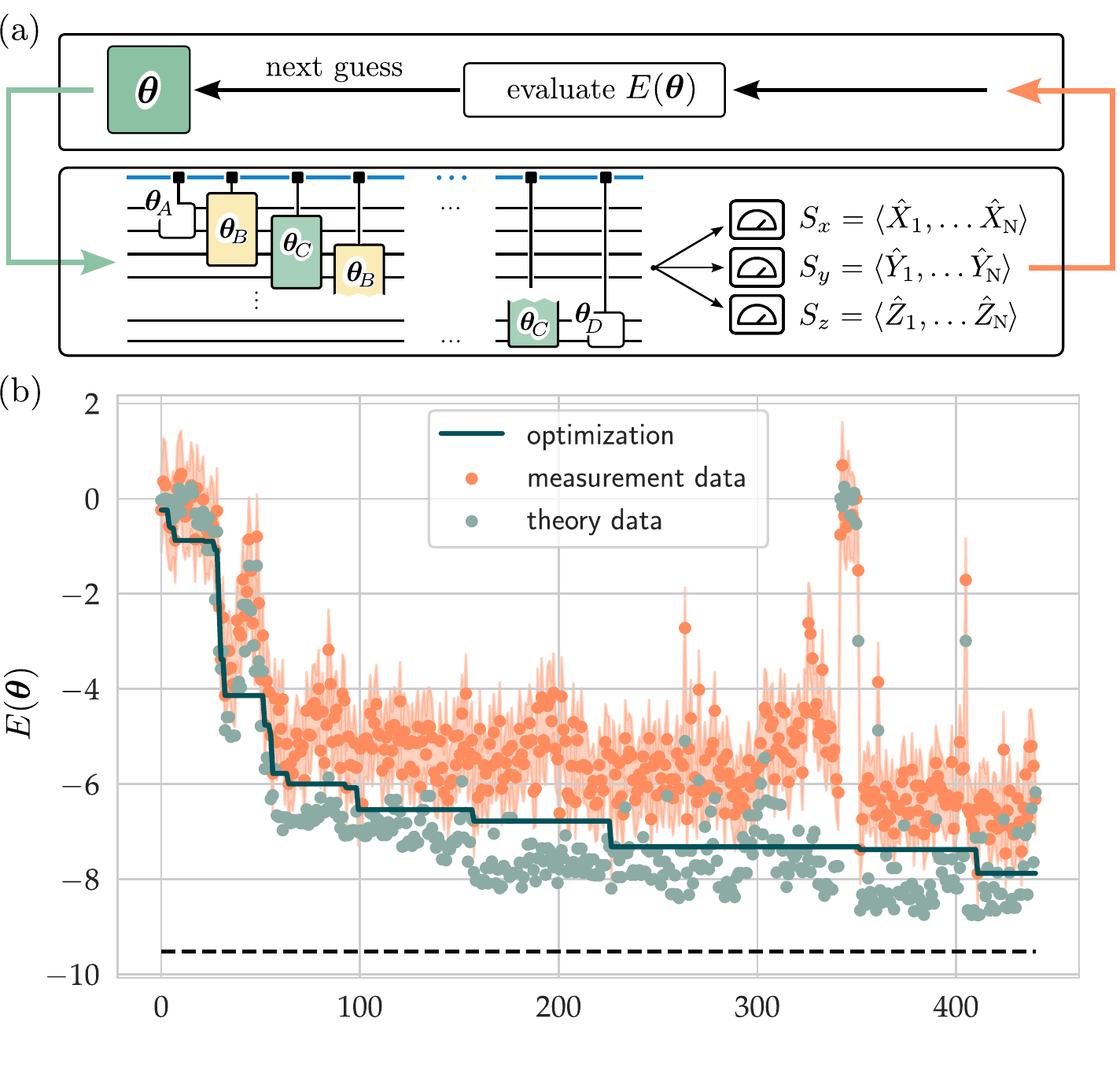}
    \caption{(a)  Sketch of the VQE protocol.  (b) Energy $E(\boldsymbol\theta)$ of the prepared variational states $\ket{\Psi(\boldsymbol\theta)}$ for $t_{-} = \delta = 0$ during closed loop optimization -- errors are calculated considering shot noise and are shown as error bars. For each step we show the immediate minimum that was reached thus far by a solid line; the energy of the exact ground state is shown as dashes. The numerical data is derived from simulations of the circuits for the respective parameters $\boldsymbol\theta$.
    }
    \label{fig:fig4_optimization_run}
\end{figure}
\noindent
The objective of our variational quantum eigensolver is to prepare the many-body ground state of the target Hamiltonian $H_{\mathrm{eSSH}}$ of Eq.~(\ref{eq:SSHmodel}) via closed-loop optimization. For a given set of couplings $t_{-}\in [-1, 1]$ and $\delta > 0$ the protocol proposes sets of trial parameters $\boldsymbol\theta$ and evaluates the energy functional
\begin{equation}
    E(\boldsymbol{\theta})=\text{Tr}\left\{ \rho_{\text{out}}(\boldsymbol{\theta})H_\mathrm{eSSH}\right\}
    \label{eq:energy_functional}
\end{equation}
via data obtained directly on the ion trap quantum simulator. A simplified sketch highlighting the full VQE is shown in Fig.~\ref{fig:fig4_optimization_run}(a). For each parameter set $\boldsymbol\theta$ we implement the circuit Fig.~\ref{fig:fig1_circuit_ions}(b) and measure collectively in the $X$, $Y$ and $Z$ basis. The measurement outcome is then fed back to the classical computer, which evaluates Eq.~(\ref{eq:energy_functional}) and makes an improved guess for new input parameters, thus iteratively minimizing $E(\boldsymbol\theta)$.
\newline

\noindent
The classical optimization algorithm that minimizes the energy functional over the parameters $\boldsymbol{\theta}$ is based on the pattern search algorithm \cite{PatternSearchHookeJeeves, PatternSearchTorczon}. This local search algorithm moves around in parameter space by polling nearby points to a candidate solution $\boldsymbol{\theta}_c$. The polling points are organized according to a stencil, centered at the candidate solution and comprised of orthogonal vectors in each of the possible search directions. Based on the experimentally measured cost function values, sampled at each of the polling points, the algorithm decides to move the stencil to a new candidate solution $\boldsymbol{\theta}_{c+1}$. If none of the polling points provided an improvement, the size of the stencil is decreased. Contrary, upon a successful energy lowering move, the stencil size is increased. The stencil is rotated such that the first polling vector is oriented along the direction of the last successful move.

Since the cost function landscape is only sampled through noisy projective measurements, some additions to the algorithm have been made. Firstly, a gaussian process model \cite{RasmussenWilliams} is fitted to the data to provide a better estimate of the cost function in the neighbourhood of the current candidate point. This is to be compared with standard gradient based algorithms, that fit a linear model to the locally obtained function values and move the candidate solution accordingly. Here instead we fit a global model, taking into account all previous measurement outcomes. A second modification for dealing with noisy cost function values is the option for the algorithm to request additional samples at already polled points, in cases where the error bars on the energy estimates are deemed too large to be able to make a good decision on where the stencil should be moved. In this refinement step, elements of optimal computational budget allocation are employed \cite{OCBA}.

We now briefly discuss details related to the implementation in the experiment. Each set of trial parameters $\boldsymbol\theta$ is defined by 14 individual angles $\theta_k$ in units of $\pi$, which are transpiled into a sequence of sideband operations adhering the proposed pattern in Fig.~\ref{fig:fig1_circuit_ions}(b) \citep{Kuzmin2020}. However, the experimental setup imposes a lower limit on the $\theta_k$ due to electronic components. Any angle yielding a laser pulse duration below $10\mathrm{\mu s}$ cannot be faithfully implemented and is thus automatically dropped by the transpiler. With a target Rabi frequency of $\Omega\approx 2\pi\cdot 8\mathrm{kHz}$ this pulse length translates to an angle $\approx 0.03\pi$. Thus, the actual circuit might differ from Fig.~\ref{fig:fig1_circuit_ions}(c). Nevertheless, simulations show that such small angles have negligible impact on the implemented quantum state.

An example of the closed-loop optimization presenting real data for the parameters $t_{-} = 0$ and $\delta = 0$ - the critical point in the thermodynamic limit - is shown in Fig.~\ref{fig:fig4_optimization_run}(b). Starting with an initial guess $\boldsymbol\theta = \lbrace 0\rbrace$ the algorithm evaluates Eq.~(\ref{eq:energy_functional}) for each optimization step. The solid line indicates the minimum for $E(\boldsymbol\theta)$ in the experimental data achieved thus far. For comparison we show the value from numerical simulations of the circuit for each of the proposed parameter sets. 
We show our `best guess' reaching $E(\boldsymbol\theta_{\mathrm{opt.}})\approx -8$ compared to the exact value of $E_\mathrm{min.} = -9.52$. 
\section{Results}
\label{sec:measurements}
\noindent
In this section we present the experimental verification of  different phases in the eSSH model by means of many-body topological invariants as phase detectors. For each data set we first run the variational optimization algorithm and, upon convergence, obtain an optimal parameter set $\boldsymbol\theta_\mathrm{opt.}$. In our experiment, the quantum state of the bulk region -- i.e.,\ the four ion state $\rho_{3, 4, 5, 6}$ in the center of the string, as shown in Fig.~\ref{fig:fig2_overview}(d) -- is measured via full state tomography and reconstructed using maximum-likelihood methods \citep{TomographyMLEHradil_1997, MLE_Banaszek}. We calculate the MBTIs $\tilde{Z}_R$ and $\tilde{Z}_T$ according to Eqs.~(\ref{eq:mbti_zr}) and (\ref{eq:mbti_zt}) from the reconstructed density matrices and calculated the measurement errors via bootstrapping. We also compare the experimental results with (i) the target values from the exact finite-size ground states, (ii) values from a ground state close to the thermodynamic limit, and (iii) values from the states obtained by numerically simulating the circuits with the experimentally obtained optimal parameter sets $\boldsymbol\theta_\mathrm{opt.}$.
\begin{figure}[tb]
    \centering
    \includegraphics[scale=0.6]{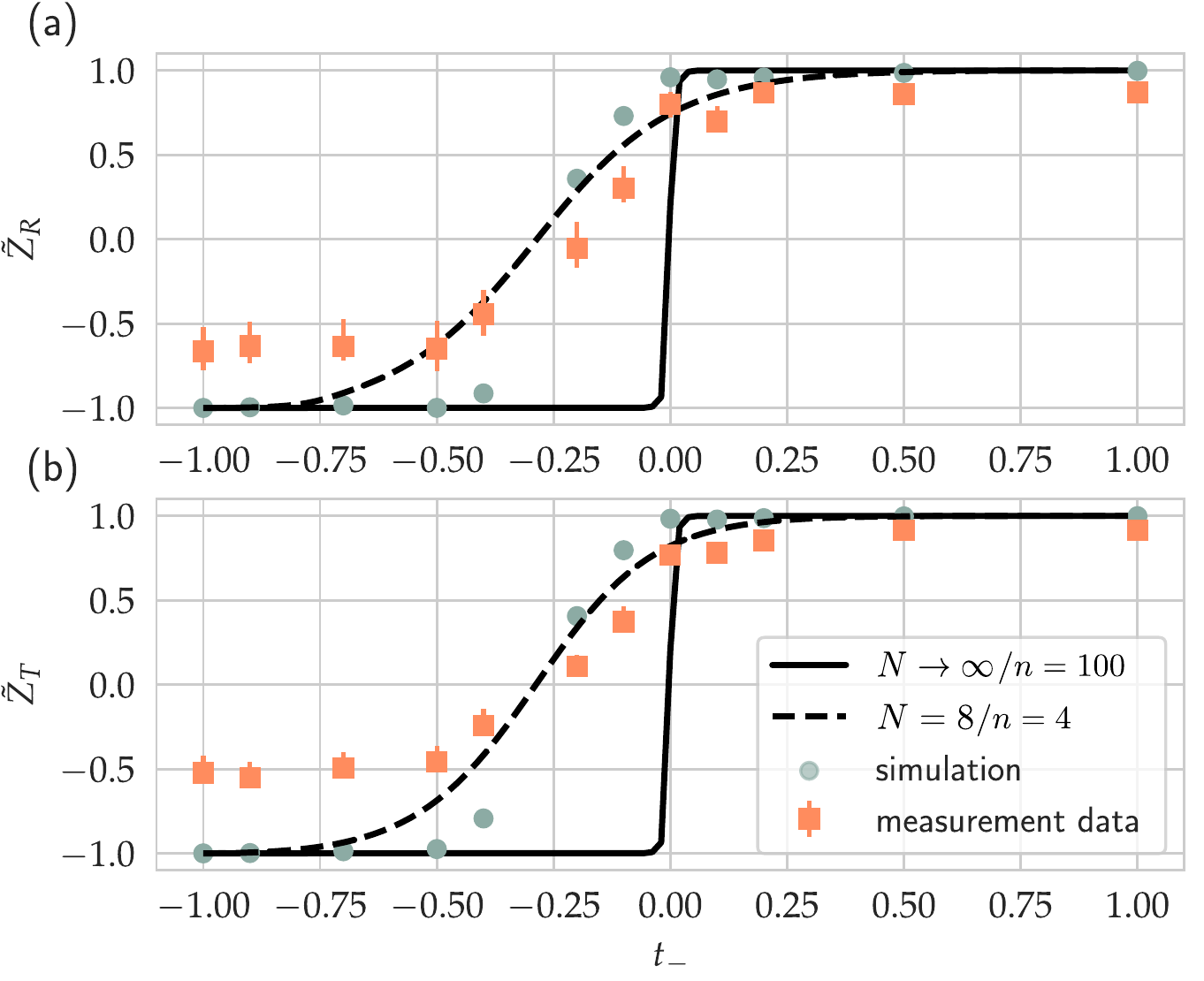}
    \caption{Many-body topological invariants $\tilde{Z}_R$ (spatial reflection) and $\tilde{Z}_T$ (time reversal) with $N=8$ and $n=4$ in the eSSH model for $\delta = 0$ as a function of $t_{-}$. For comparison, a simulation of the input parameters $\boldsymbol{\theta}_{\mathrm{opt.}}$ is included together with $\tilde{Z}_{R/T}$ for the ground states in the thermodynamic limit $N\to\infty$, found by iDMRG, and the finite system with $N=8$.}
    \label{fig:fig5_mbti_delta_0}
\end{figure}
\begin{figure}[tb]
    \centering
    \includegraphics[scale=0.6]{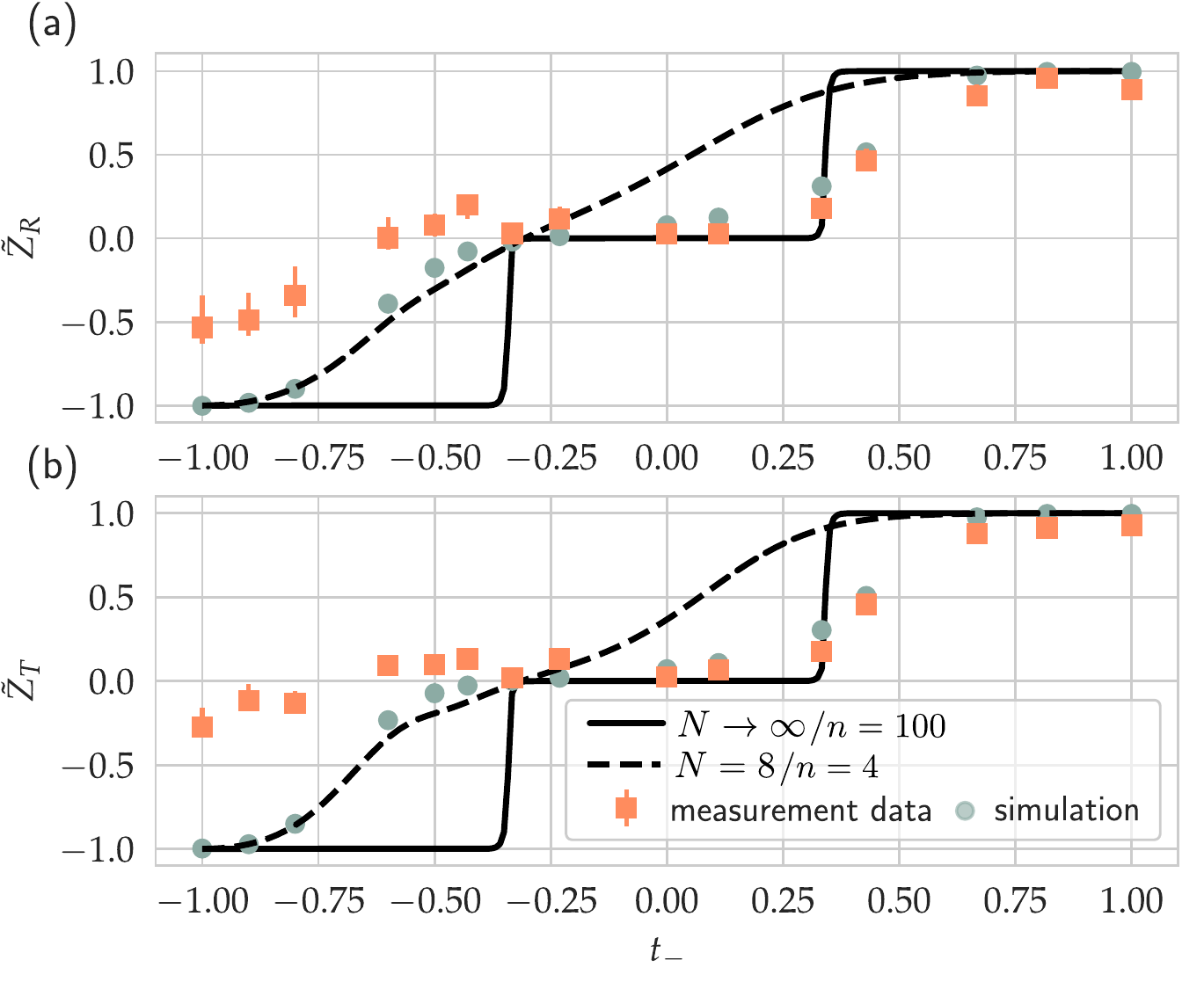}
    \caption{MBTI $\tilde{Z}_R$ and $\tilde{Z}_T$ in the eSSH model for $\delta=4$, indicating a symmetry-broken phase in the region near $t_{-}\approx 0$. The notations are the same as in Fig.~\ref{fig:fig5_mbti_delta_0}}
\label{fig:fig6_mbti_delta_4}
\end{figure}

First we discuss the case $\delta = 0$, where Eq.~(\ref{eq:SSHmodel}) is equivalent to the plain SSH model. In the thermodynamic limit, the transition between the topological and trivial phases occurs with increasing $t_-$ at $t_-=0$ and is indicated by the abrupt change of the MBTIs values from $-1$ to $1$, as given by solid black lines in Fig.~\ref{fig:fig5_mbti_delta_0}. However, in the finite-size ground states, the MBTIs, shown by dashed black lines, change smoothly. Moreover, the center of the slope is shifted from $t_-=0$ to the negative values due to finite-size effects, and specifically, that the number of odd-even terms in the eSSH Hamiltonian, Eq~(\ref{eq:SSHmodel}), dominates the number of even-odd terms by one.

The data obtained in our experiment shown by orange square markers is in good agreement with the described behavior and clearly demonstrates the transition of the MBTIs from the negative values in the topological phase to the positive in the trivial phases. We observe the most pronounced deviation of the experimental values from the target values and from the numerical simulation in the topological phase, especially with respect to the time-reversal symmetry in Fig.~\ref{fig:fig5_mbti_delta_0}(b). This observation is attributed to decoherence processes in the COM mode and residual Stark shifts, resulting in depolarization of the prepared states along the quantization axis of the qubits, which both MBTIs are sensitive to. A more detailed simulation analysis of this qualitative argument is included in Appendix~\ref{app:app_error_modelling}. Considering the example states shown in Fig.~\ref{fig:fig2_overview}(d), it can be shown that both the MBTIs are quite fragile, under dephasing, in the topological phase, while they are relatively robust in the trivial phase. This effect is due to the algebraic properties of the MBTI itself, given that we consider a bulk of four sites. In fact, in the topological phase, losing coherences can cause $\tilde{Z}_{R}$ to increase all the way up to zero. Conversely, in the trivial phase, $\tilde{Z}_{R}$ can decrease only to $1/\sqrt{2}$. Such asymmetry is reflected in the quality of the MBTIs in either phase.

Having explored the integrable case, we now consider the case $\delta = 4$, which exhibits the symmetry-broken phase (an Ising Antiferromagnet) between the topological and trivial phases due to the sharp anisotropy favoring interactions in the $z$-direction. This phase is indicated by the abrupt change of the MBTIs to $0$ value, as shown in Fig.~\ref{fig:fig6_mbti_delta_4}. However, the exact finite-size ground state indicates almost no presence of the symmetry-broken phase. Interestingly, the symmetry-broken phase can be clearly observed from the experimental data. We attribute this behavior to the fact that at a finite size, the exact model resolves the ground state degeneracy, and, in the finite-size unique ground state, the parity symmetry is not spontaneously broken. In contrast, due to imperfections and noise, experimental VQE polarizes into a physical ground state with spontaneous symmetry breaking, see middle panel in Fig.~\ref{fig:fig2_overview}(d). This state is less entangled, has smaller finite-size effects, and is thus closer to the values of the thermodynamic limit. For more details, see Appendix~\ref{app:app_error_modelling}. Similar to the case $\delta = 0$, we also observe deviation of the experimental values from the target values in the topological phase.

\section{Conclusions and Outlook}
\label{sec:conclusions_and_outlook}
\noindent
In the present work, we implemented a variational quantum eigensolver in an ion trap quantum device, capable of targeting tensor network states. We demonstrated that our technique is efficient at preparing entangled ground states of gapped Hamiltonians, including symmetry-protected topological phases.
Our strategy encodes the tensor network variational ansatz in  a quantum circuit which explicitly includes the COM vibrational mode as an entanglement mediator. The native interactions of the ions with the COM mode were used in our circuit as variational resource operations, and the target pure state was approximated in ions via a variational quantum eigensolver. We carried out our experiments in traps with 8 ions, and successfully prepared each gapped phase of the non-integrable interacting extension of the SSH model at zero temperature.

We view our work as a first step to realize a scalable tensor-network simulator on an ion trap platform. The quality of our experiment has been improved by spectroscopic decoupling to suppress cross-talk between the ions -- one of the major imperfections in ion traps. We observed that one of the limitations on the system size $N$ in our circuit is finite coherence time $\tau_\mathrm{COM}$ of the COM mode. Since the coherence time of the COM mode scales roughly as $1/N$ and the depth of the TN circuit is proportional to $N$, one can estimate the size limit to $N\propto\sqrt{\tau_\mathrm{COM}}$. Scalable simulation can be achieved with several sideband sequences separated by intermediate recooling of the phonon mode. In such a scheme, the ions must be variationally disentangled from the phonon mode at the end of each sequence. Additionally, implementing ion traps exhibiting local phonon modes will also circumvent this restriction \cite{osti_1237003,Tobias2020}. Our TN circuit employed translation invariance in the bulk, thus we expect that the complexity of the optimization problem will not increase with the system size but only with the entanglement growth in the target state. Another challenge for large-scale simulation of many-body phases is their verification. In ion traps, this problem can be tackled by adapting randomized measurement techniques~\cite{brydges2019probing,ElbenTopoInvariants,carrasco2021theoretical,EntanglementRandomeas}.

A crucial step towards a useful quantum simulation is the extension of the TN-VQE towards 2D lattice systems~\cite{banulsSequentiallyGeneratedStates,Soejima2019,slatteryQuantumCircuitsTwoDimensional2021,wei2021sequential} and tensor network geometries for 2D, such as Projected Entangled-Pair States (PEPS) \cite{PEPS2008,PEPS2021}.
Precisely, recent work~\cite{wei2021sequential} discusses quantum circuits for sequential generation of plaquette-PEPS -- a subclass of PEPS, that is believed to include a large class of 2D phases, including the topological ones~\cite{Soejima2019}. In Appendix \ref{app:2Dproposal}, we speculate on a potential implementation of 2D TN states in scalable ion trap architectures that allow ion-crystal reconfiguration. In particular, we discuss a detailed schedule of operations for a 5x5 lattice in a well-tested microstructured ion trap~\citep{osti_1237003}.
Alternatively, we are confident that scalable realization of the TN circuit for 2D systems in simpler trap architectures can be achieved  by implementing in-sequence projective measurements and reset of individual ions~\citep{Recycling2012}. Indeed, assessing the optimal implementation scheme for a given trap architecture will require further research.

The demonstrated experimental capabilities open opportunities for ion trap implementation of a variety of protocols that make use of 1D variational tensor-network circuits. These include a protocol to study infinite 1D systems~\cite{barrattParallelQuantumSimulation2021}, imaginary and real time evolution~\cite{linRealImaginaryTimeEvolution2021}, and quantum machine learning~\cite{huggins2019towards}. Finally, the resonant interactions with one or several phonon modes potentially can be used to construct circuits beyond the TN ans{\"a}tze, e.g., to address problems in quantum chemistry~\cite{kandala2017hardware, hempel2018quantum}. The design of the appropriate variational circuit might be obtained in a closed-loop optimization on a quantum device itself using recent hybrid algorithms such as adaptive algorithm~\cite{tang2021qubit} or reinforcement learning~\cite{ostaszewski2021reinforcement}.

\section{Acknowledgements}
\noindent
We gratefully acknowledge funding by
the Austrian Research Promotion Agency (FFG) contract 872766 (project AutomatiQ) and contract 884471 (project ELQO),
the Austrian Science Fund (FWF) via the SFB BeyondC project No. F7109,
and
the US Air Force Office of Scientific Research (AFOSR) IOE Grant No. FA9550-19-1-7044.
We further received support from the European Union's Horizon 2020 research and innovation program under the Marie Skłodowska-Curie grant agreement No. 840450, and from the IQI GMBH. Our work only reflects the views of its authors, all agencies are not responsible for any use that may be the direct or indirect result of information contained in this paper.
\appendix
\section{Analysis of experimental imperfections}
\label{app:app_error_modelling}
\noindent
In this section, we examine and compare different error models to explain the deviation between the experimental data and the predicted values for $\tilde{Z}_{R/T}$ in Fig.~\ref{fig:fig2_overview}, Fig.~\ref{fig:fig5_mbti_delta_0} and Fig.~\ref{fig:fig6_mbti_delta_4} of the main text. We numerically simulate the variational circuit for $\delta = 4$ with the experimentally obtained optimal parameter sets $\boldsymbol\theta_{\mathrm{opt.}} $, considering the full circuit as a sequence of operations on the blue sideband -- errors are modelled to occur after each individual sideband operation. We quantify the performance of the considered error models via the residual sum of squares (RSS)
\begin{equation*}
    \mathrm{RSS} = \sum_i \left(\tilde{Z}_i^{\mathrm{data}} - \tilde{Z}_i^{\mathrm{model}}\right)^2
\end{equation*}
\noindent
with $i$ labelling each point obtained for $\tilde{Z}_{R/T}$. From the following analysis, we conclude that, in our task, the initial temperature of the COM mode can significantly affect only the symmetry-broken phase. Because of the finite temperature, VQE can prepare a low-energy symmetry-broken state instead of the symmetric exact ground state. However, this imperfection is not relevant for investigating condensed matter phases. We identified the dominant sources of errors as the COM mode heating, fluctuations of the tip voltages of the trap electrodes, and imperfect compensation of the Stark shift, which effectively result in depolarization of the prepared states along the quantization axis of the qubits. Particularly, the coherence time of the COM mode should scales roughly as $1/N$ with $N$ the number of ions. Since the depth of our variational circuit is proportional to $N$, one can estimate the size limitation as $N\propto\sqrt{\tau_\mathrm{COM}}$.
\vspace{-10pt}
\subsubsection{Finite temperature of the COM mode}
\begin{figure}[tb]
    \centering
    \includegraphics[width=1\columnwidth]{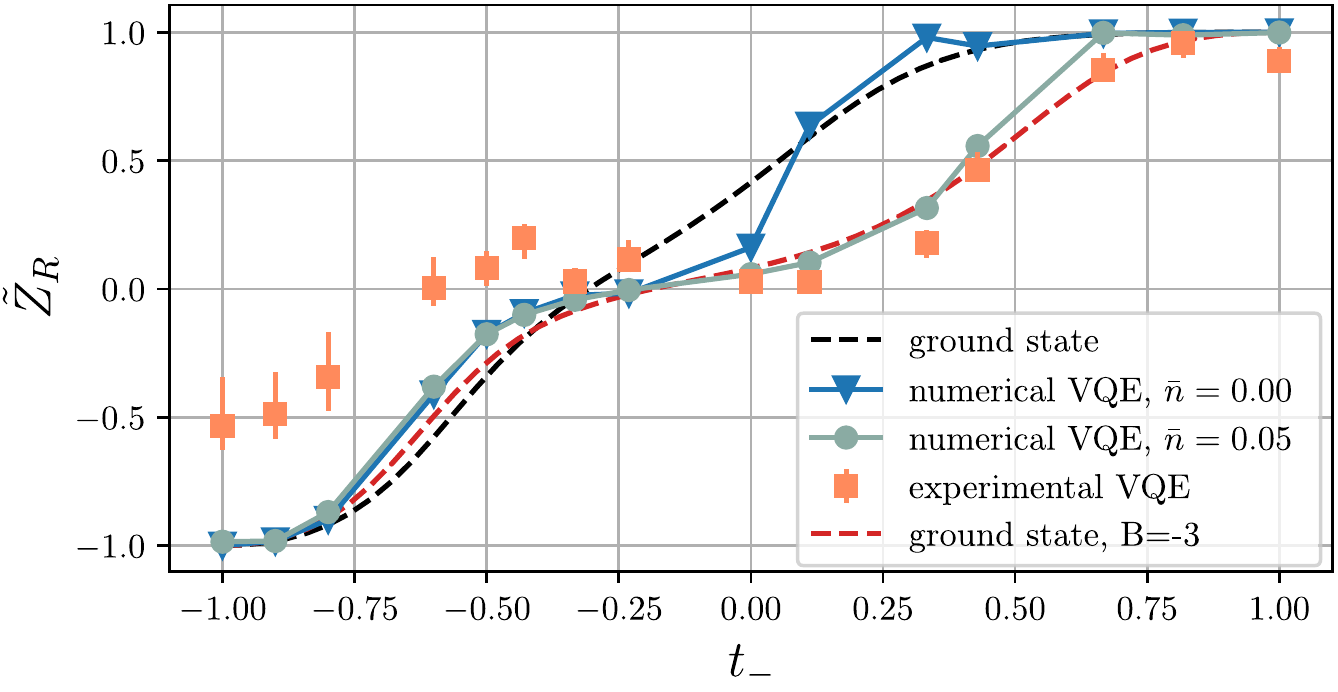}
    \caption{Partial reflection MBTI $\tilde{Z}_R$, Eq.~(\ref{eq:mbti_zr}), for the extended SSH model, Eq.~(\ref{eq:SSHmodel_with_B}), at $\delta=4$ with $N=8$ and $n=4$. The dashed lines give the MBTIs from the exact ground states; in black we show the ground state without a pinned magnetic field $B=0$, while red depicts the case with $B=-3$, see the text. Numerical simulations of the VQE with the COM mode having a mean phonon number $\bar{n} = 0$ and $\bar{n}=0.05$ are shown by the solid lines. The experimentally measured data is represented by squares.}
    \label{fig:appendix_Zr}
\end{figure}
\begin{figure}[tb]
    \centering
    \includegraphics[width=1\columnwidth]{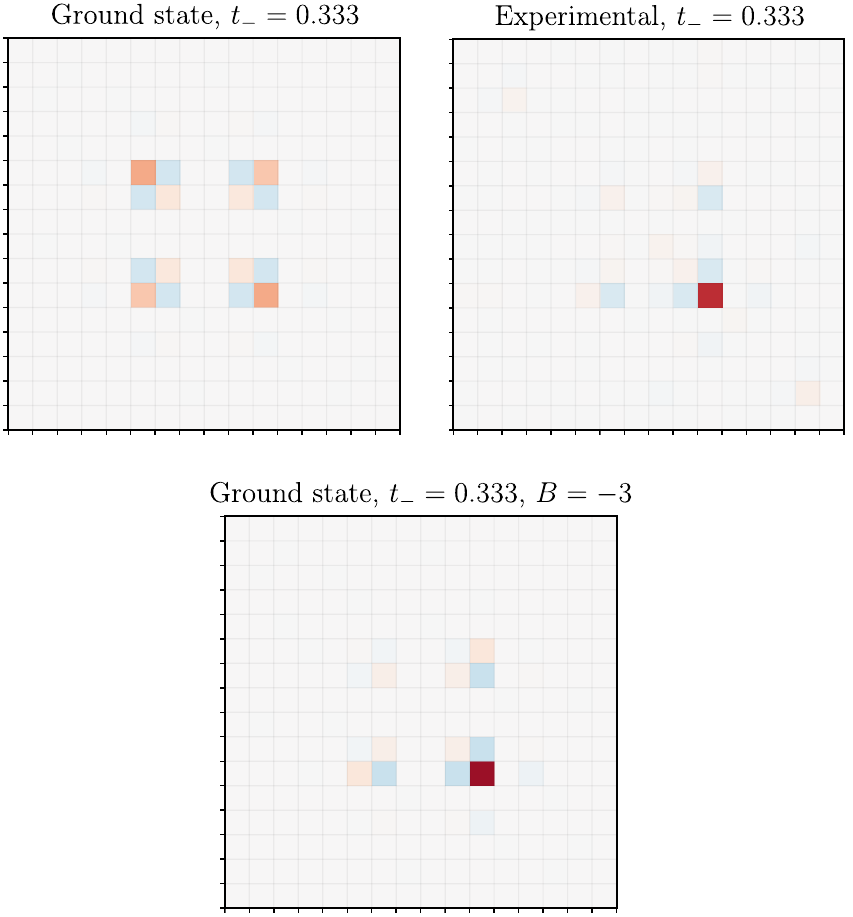}
    \caption{Reduced density matrices of the four central ions in an 8 ion system. The top left state corresponds to the ground state at $t_{-} = 0.333$. The experimentally measured data for same $t_{-}$ is shown in the right top state. In the bottom panel we show the ground state of the eSSH model with a pinned magnetic field $B = -3$, see Eq.~(\ref{eq:SSHmodel_with_B}), for $t_{-} = 0.333$, which is in good agreement with the experimental data. }
    \label{fig:appendix_dms}
\end{figure}
\begin{figure}[tb]
    \centering
    \includegraphics[width=1\columnwidth]{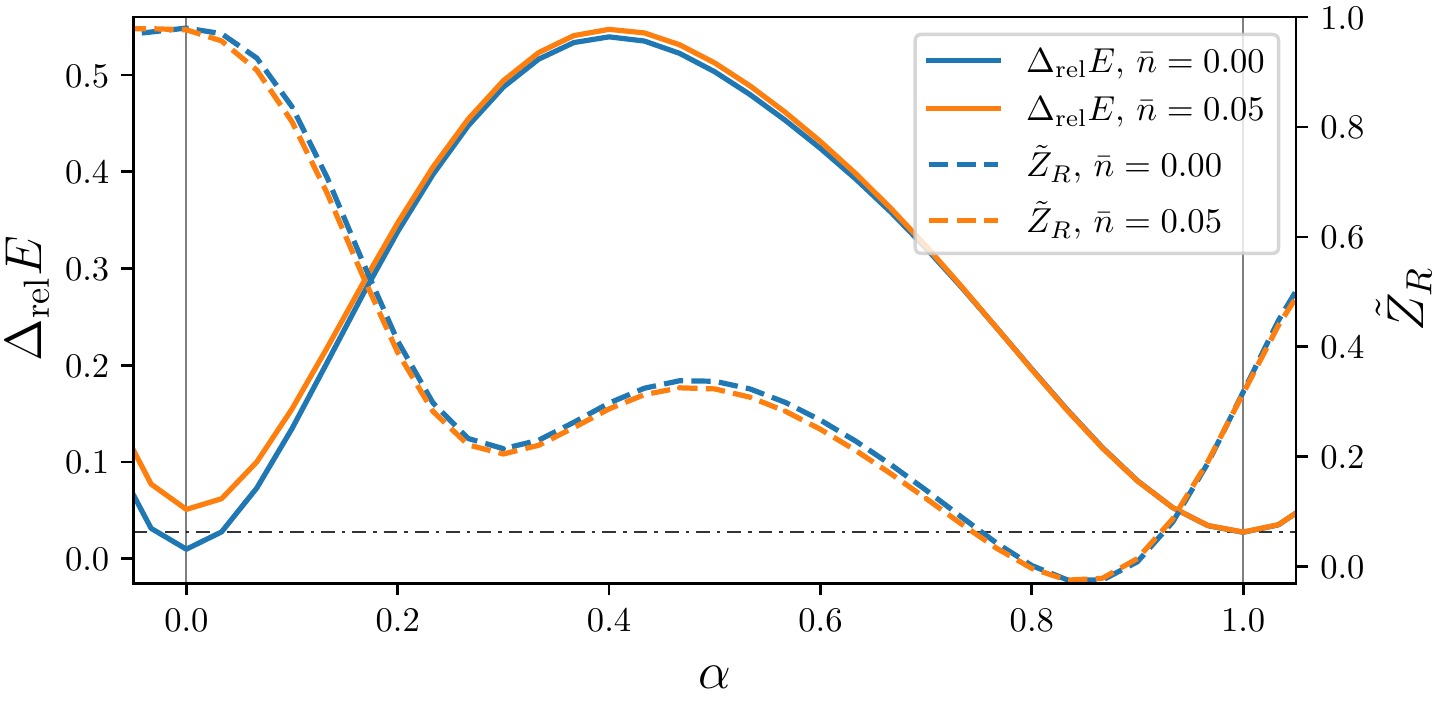}
    \caption{Numerical simulation of the ansatz circuit in Fig.~\ref{fig:fig1_circuit_ions} for a range of parameter sets $\boldsymbol\theta(\alpha)$ parametrized by $\alpha$, see the text, for the mean phonon numbers $\bar{n} = 0$ and $\bar{n} = 0.05$. The left vertical axis gives the relative energy error of the obtained states with respect to the exact ground state energy of the eSSH model at $\delta = 4$, $t_{-} = 0.333$; the right axis gives the corresponding value of partial reflection MBTI $\tilde{Z}_R$.}
    \label{fig:appendix_E_rel}
\end{figure}
\noindent
We investigate the effect of finite temperature of the COM mode in the initial state caused by imperfect cooling. In Fig.~\ref{fig:appendix_Zr} we show the results of the numerical simulation of the VQE where we consider the mean phonon numbers $\bar{n} = 0.00$ and $\bar{n} = 0.05$ in the thermal state of the COM mode. Here we do not include heating and dephasing, as these mechanisms are considered individually in the next section. For both values of $\bar{n}$ we observe no shift of $\tilde{Z}_R$ in the topological phase, which, in contrast, is present in our experimental data. However, in the symmetry-broken phase, unlike for $\bar{n} = 0$, the case for $\bar{n} = 0.05$ demonstrates results close to the experimentally obtained values. For $\bar{n} = 0.05$  we obtain a residual sum of squares $\mathrm{RSS} = 1.11(96)$ for $\tilde{Z}_R$ and $\mathrm{RSS} = 1.92(48)$ for $\mathrm{Z}_T$ -- when averaged, we get $\mathrm{RSS} = 1.52(22)$). In contrast, $\bar{n} = 0.00$ yields larger deviations, namely $\mathrm{RSS} = 2.41(76)$ for $\tilde{Z}_R$ and $\mathrm{RSS} = 3.23(46)$ for $\mathrm{Z}_T$ (averaged $\mathrm{RSS} = 2.82(61)$).

We attribute the significant difference in the MBTIs for the cases $\bar{n} = 0.00$ and $\bar{n} = 0.05$ in the symmetry-broken phase to the ground state degeneracy. At finite size, the exact model resolves this degeneracy. In the unique finite-size ground state the parity symmetry is not spontaneously broken, as shown at the top left panel of Fig.~\ref{fig:appendix_dms}, which gives the reduced states of 4 middle qubits in the ground state of the eSSH model at $\delta=4$ and $t_-=0.333$. In contrast, due to the finite temperature of the COM mode, the VQE algorithm polarizes into a physical ground state with spontaneous symmetry breaking as indicated by the top right panel in Fig.~\ref{fig:appendix_dms}. This state is less entangled, has smaller finite-size effects and is thus closer to the value $\tilde{Z}_R=0$ in the thermodynamic limit. 

The fragility of the global energy minimum corresponding to the exact ground state of the eSSH model at $\delta = 4$, $t_{-} = 0.333$ is shown in Fig.~(\ref{fig:appendix_Zr}). We simulated the variational circuit for a range of parameters $\boldsymbol\theta(\alpha)=(1-\alpha)\boldsymbol\theta_\mathrm{opt.}|_{\bar n = 0}+\alpha\boldsymbol\theta_\mathrm{opt.}|_{\bar n = 0.05}$, which interpolates between the optimal parameters $\boldsymbol\theta_\mathrm{opt.}|_{\bar n = 0}$ found by VQE for the cases with $\bar n = 0$ and the optimal parameters $\boldsymbol\theta_\mathrm{opt.}|_{\bar n = 0.05}$ found for $\bar n = 0.05$.
For the respective states we calculate the relative energy error $\Delta_\mathrm{rel}=[E_\mathrm{targ}-E(\boldsymbol\theta)]/E_\mathrm{targ}$ with respect to the ground state energy and compute the corresponding values of $\tilde{Z}_R$. At $\alpha=0$ and $\alpha=1$ $\Delta_\mathrm{rel}$ exhibits two local minima. For $\bar n = 0$ the global minimum corresponds to the symmetric state at $\alpha=0$, while at $\bar n = 0.05$ we observe a substantial energy shift. In contrast, the symmetry-broken state at $\alpha=1$ demonstrates almost no energy shift when $\bar n$ increases, thus its energy becomes the global minimum for $\bar n = 0.05$. This robustness is a result of the smaller amount of entanglement in the reduced bulk state compared to the exact ground state -- this is highlighted by the top panels in Fig.~\ref{fig:appendix_dms}).

Breaking of the symmetry can also be achieved in the target eSSH model by introducing sufficiently large pinned staggered magnetic field $B$ on the outermost ions
\begin{multline}
    H_{\text{eSSH}}=\sum_{k=1}^{N-1}\left[1+(-1)^{k-1}t_{-}\right]\\
    \times\left(\sigma_{k}^{x}\sigma_{k+1}^{x}+\sigma_{k}^{y}\sigma_{k+1}^{y}+\delta\sigma_{k}^{z}\sigma_{k+1}^{z}\right)+B\left(\sigma_{1}^{z}-\sigma_{N}^{z}\right).
    \label{eq:SSHmodel_with_B}
\end{multline}
For $B=-3$ the exact ground state is in good agreement with the experimental data as shown in Fig.~\ref{fig:appendix_Zr} by the red dashed line -- this is also evident from the density matrices in Fig.~\ref{fig:appendix_dms} when directly comparing the top right and bottom panels.  With increasing system size, the gap between the ground state and the first excited states decreases, in turn also decreasing the required pinning field $B$ until it vanishes in the thermodynamic limit.

Our analysis shows, that for fixed circuit parameters $\boldsymbol\theta$ moderate temperature of the initial state of the COM mode does not affect the MBTIs significantly. Instead it can cause the VQE to obtain different optimal parameters $\boldsymbol\theta_{\mathrm{opt}}$, preparing a physical low-energy state with broken symmetry instead of the exact symmetric ground state in the symmetry-broken phase. This imperfection, however, is not important when investigating phases of quantum matter, since -- in the thermodynamic limit -- the same symmetry breaking occurs spontaneously regardless.
\vspace{-10pt}
\subsubsection{Heating of motional modes}
\noindent
Residual noise, most importantly due to electric field fluctuations in the ion trap, disturbs the motional state of the ion string, causing the string to heat up. In our setup, these perturbations occur at a rate $\Gamma_{\mathrm{H}} = 27(1)$ phonons per second. This effect is modelled by the channel 
\begin{equation*}
    \rho \to \rho' = (1 - p)\cdot\rho + p\cdot \hat{a}^\dagger\varrho\hat{a},
\end{equation*}
inducing an error with probability $p = t\Gamma_{\mathrm{H}}$ on the state $\rho$ depending on the time $t\ll \nicefrac{1}{\Gamma_{\mathrm{H}}}$. We simulate heating in the circuits with the initial states having different temperatures, namely $\bar{n} = 0.00$ and $\bar{n} = 0.05$. To establish a notion of time, we assume each blue sideband operation implementing $\ket{S, n=0}\to\ket{D, n=1}$ to require a laser pulse of length $t_\pi =\SI{125}{\mu s}$. 
\begin{figure}
    \centering
    \includegraphics[scale=0.58]{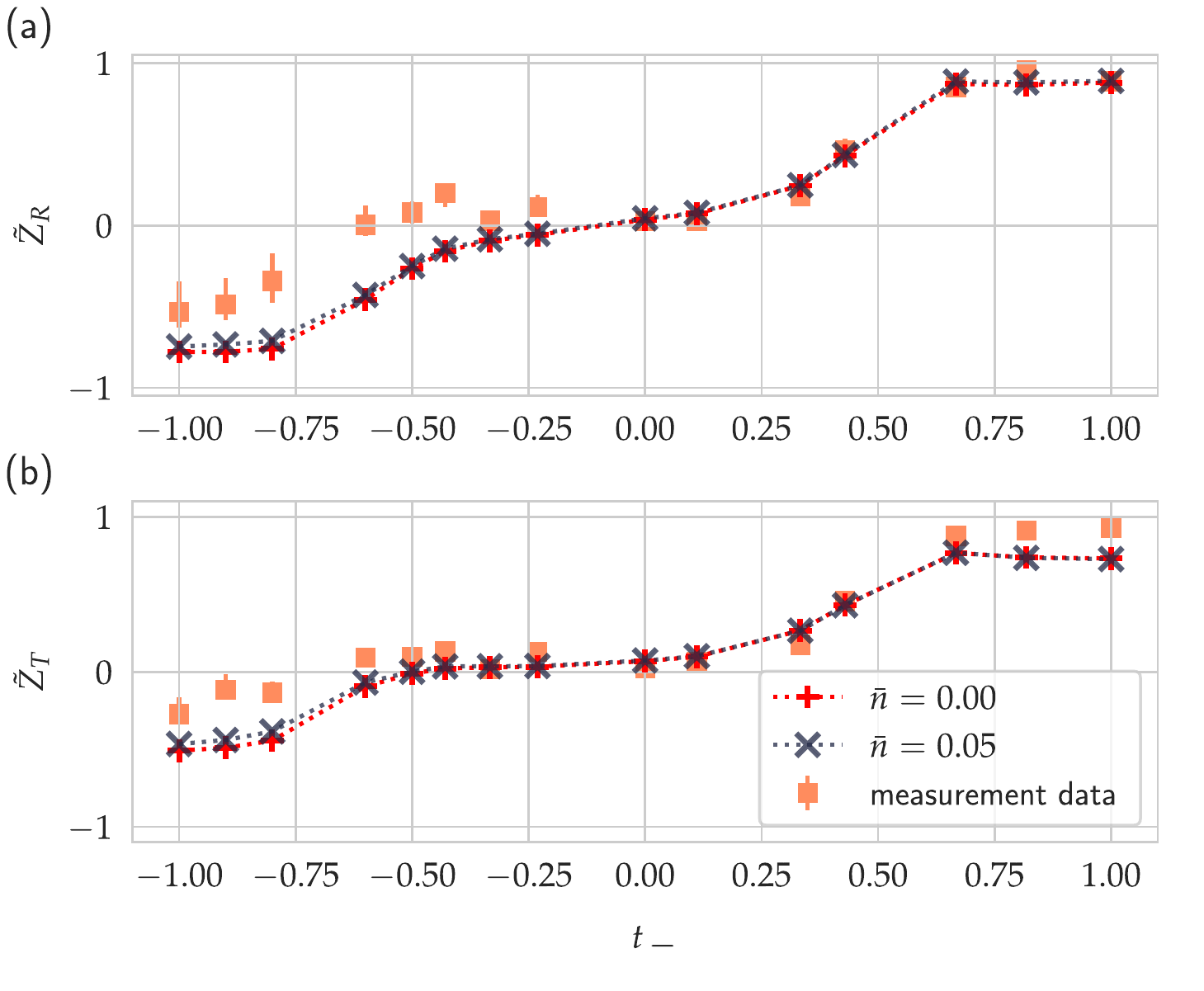}
    \caption{Simulated heating of the motional modes for mean phonon numbers $\bar{n} = 0.00$ and $\bar{n} = 0.05$ of the initial state. The time reference is established via the average laser pulse length $t_\pi = \SI{125}{\mu s}$ required to implement a $\pi$-flip on the blue sideband.}
    \label{fig:app_mbti_errors_heating}
\end{figure}
In Fig.~\ref{fig:app_mbti_errors_heating}(a/b) we observe a significant effect on the MBTIs $\tilde{Z}_{R/T}$ in the topological phase $t_{-}\to -1$, which is less pronounced in the trivial phase $t_{-}\to +1$. However, we see deviations in $\tilde{Z}_T$, which tend to increase as we approach $t_{-} = 1$. These can be attributed to different lengths $T$ required to execute the pulse sequences as shown in Fig.~\ref{fig:app_mbti_sequence_lengths}. As $T$ grows when approaching the edge cases $t_{-}\to \pm 1$, these sequences are more affected by heating effects. Considering a perfectly cooled initial state with a mean phonon number $\bar{n}=0$, the residual sum of squares are given by $\mathrm{RSS} = 0.84(21)$ for $\tilde{Z}_R$ and $\mathrm{RSS} = 0.45(11)$ for $\tilde{Z}_T$ - when averaged, we obtain $\mathrm{RSS} = 0.65(16)$. Increasing the initial temperature to $\bar{n} = 0.05$ yields $\mathrm{RSS} = 0.71(18)$ for $\tilde{Z}_R$ and $\mathrm{RSS} = 0.35(10)$ for $\mathrm{Z}_T$ (average $\mathrm{RSS} = 0.65(16)$). However, as discussed in the previous subsection, the temperature of the initial state does not affect the MBTIs significantly for fixed circuit parameters.
%s
\begin{figure}[b]
    \centering
    \includegraphics[scale=0.58]{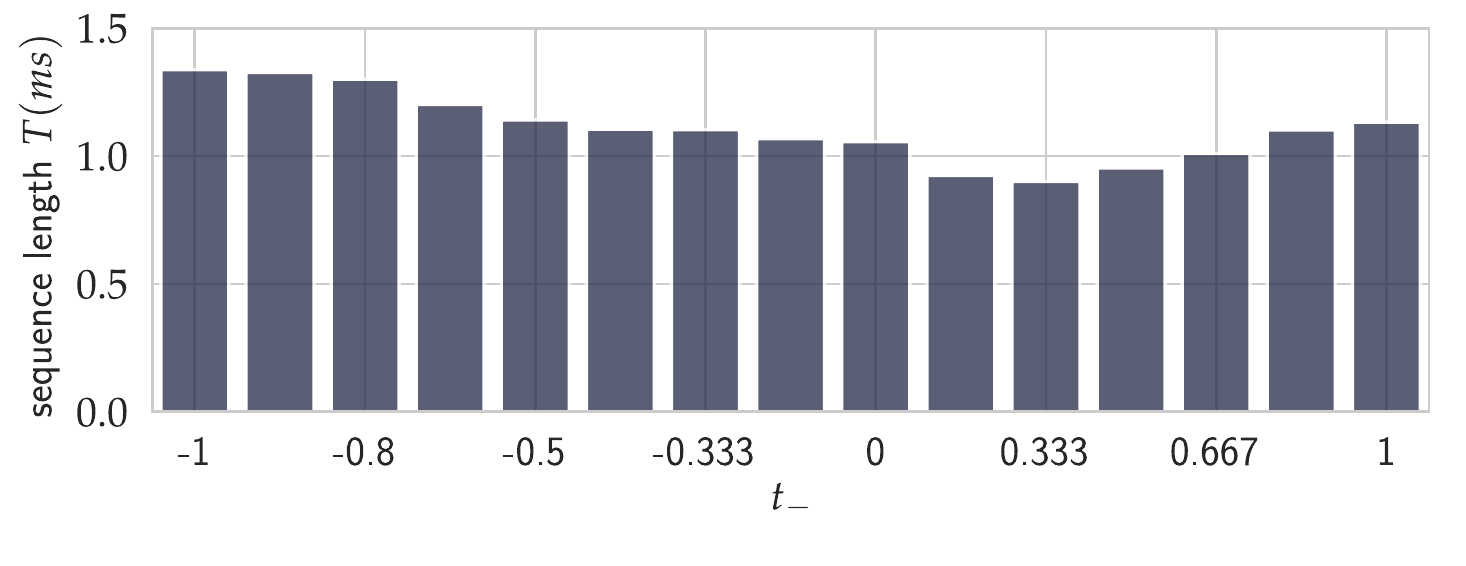}
    \caption{Lengths $T$ of the individual laser pulse sequences required to implement the desired target state. $T$ increases as $t_{-}$ approaches $\pm 1$, giving the system more time to be affected by heating.}
    \label{fig:app_mbti_sequence_lengths}
\end{figure}
\subsubsection{Depolariziation \& weighted Pauli errors}
\noindent
Decoherence processes on the motional modes translate to errors on the addressed qubits. Such errors can be modelled by a depolarizing channel, where each spatial axis is equally affected by a product of single qubit Pauli errors $\sigma_i^{(x,y,z)}$, with probability $\nicefrac{p}{3}$. In reality, it is unlikely that all errors are distributed equally -- thus, we must consider probabilities for each axis $p_x$, $p_y$ and $p_z$, such that for a system of $N$ qubits
\begin{equation*}
    \rho' = (1 - \hspace{-1em}\sum_{j \in\{x,y,z\}}\hspace{-1em} p_j)^N\rho + \prod_i^N \sum_{j \in\{x,y,z\}} \hspace{-1em} p_j\hat{\sigma}_i^j.
\end{equation*}
For simplicity we assume $p_x = p_y = p_{xy}$ -- this assumption was shown to be valid in simulations, where exchanging $p_x$ for $p_y$ and vice versa yields identical results. We evaluate the model for both $\tilde{Z}_R$ and $\tilde{Z}_T$ for all combinations $p_{xy, z} = \left\lbrace 0, 0.001, 0.005, 0.01, 0.02, 0.03, 0.04\right\rbrace$ and estimate the `goodness of fit` via the RSS to deduce the most likely source of error in our setup. Note, that this analysis also considers the case without any errors, i.e. $p_{xy} = p_z = 0$. 

\begin{figure}[h!]
    \centering
    \includegraphics[scale=0.54]{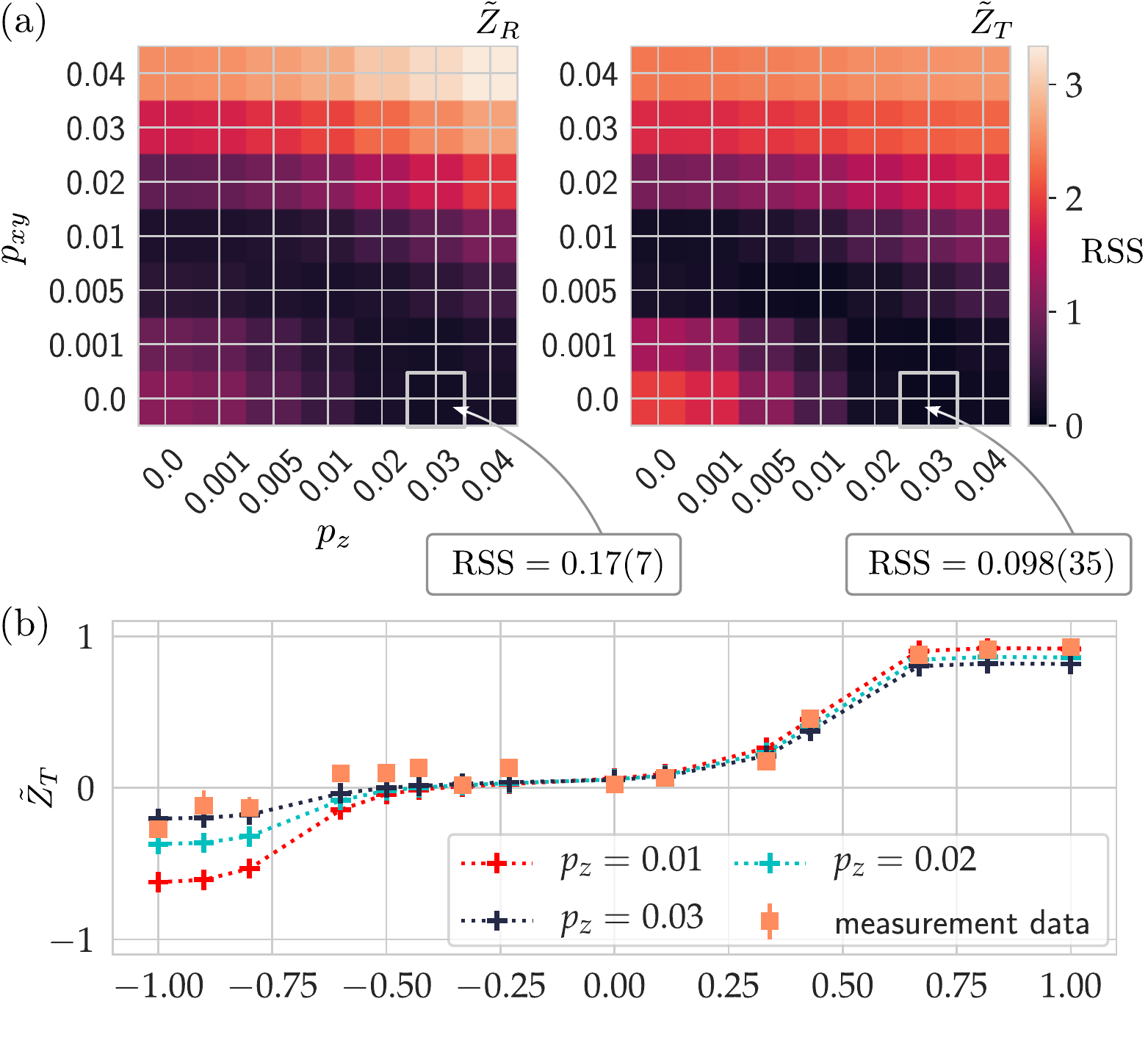}
    \caption{`Goodness of fit` estimation via residual sum of squares for the weighted Pauli error model. We compare the model with the data for $\tilde{Z}_R$ and $\tilde{Z}_T$ individually in (a). As we expect $X$ and $Y$ errors to occur with the same probability, we assume $p_x = p_y = p_{xy}$. The best fit is achieved for $p_{xy} = 0$ and $p_z = 0.03$ with $\mathrm{RSS} = 0.17(7)$ in $\tilde{Z}_R$ and $\mathrm{RSS} = 0.098(35)$ in $\tilde{Z}_T$ shown in (b), respectively.}
    \label{fig:app_model_deviation_rss}
\end{figure}

\begin{figure}[h!]
    \centering
    \includegraphics[scale=0.54]{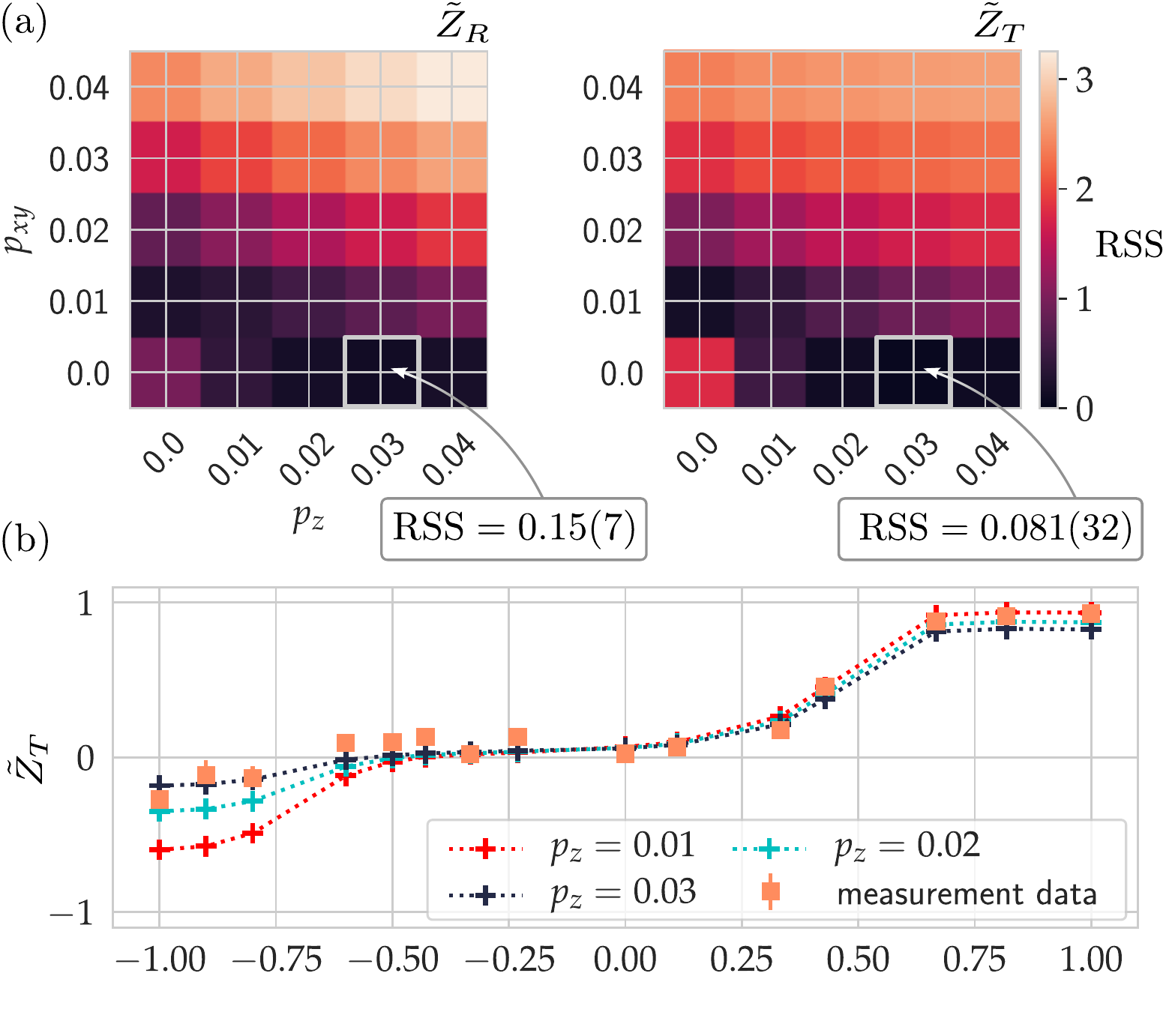}
    \caption{(a) Residual sum of squares of the weighted Pauli error model for the MBTIs $\tilde{Z}_R/T$ considering an imperfectly cooled initial state with mean phonon number $\bar{n} = 0.05$ for different probabilities $p_{xy,z}$. We obtain an $\mathrm{RSS} = 0.15(7)$ in $\tilde{Z}_R$ and $\mathrm{RSS} = 0.081(32)$ (average $\mathrm{RSS} = 0.12(5)$) shown in (b) for $\tilde{Z}_T$.}
    \label{fig:app_model_deviation_rss_warm}
\end{figure}

The results of our comparison are shown in Fig.~\ref{fig:app_model_deviation_rss}(a/b) -- the best fit is achieved, when $p_{xy} = 0$ and $p_z = 0.03$ for both MBTIs. We obtain a $\mathrm{RSS} = 0.17(7)$ for $\tilde{Z}_R$ and $\mathrm{RSS} = 0.098(35)$ in $\tilde{Z}_T$ ($\mathrm{RSS} = 0.13(5)$ when averaged) and conclude, that the circuit is predominantly impacted by dephasing, resulting in $Z$-type errors. From a purely experimental point of view, this argument is expected for two reasons: first, even in the presence of laser intensity and laser phase noise we are able to control the rotation angles $\theta$ and $\phi$ of single qubit gates to an extend, such that we can achieve average gate fidelities of beyond $99.99\%$. Second, the VQE algorithm is supposed to take systematic over- and underrotations into account when converging to an optimal set of circuit parameters $\boldsymbol\theta_{\mathrm{opt.}}$. The observed $Z$-errors are a consequence of various sources:  we expect contributions due to heating and dephasing of the motional states as well as residual detuning from the first blue sideband, which is a consequence of imperfect Stark shift compensation. 
\newline\newline\noindent
We also consider a finite temperature of the initial state with mean phonon number of $\bar{n} = 0.05$ and evaluate the models performance. The results are shown in Fig.~\ref{fig:app_model_deviation_rss_warm}; we find a best fit for the probabilities $p_{xy} = 0$ and $p_z = 0.03$ in agreement with the data obtained for the ideally cooled initial state. The residual sum of squares is $\mathrm{RSS} = 0.15(7)$ for $\tilde{Z}_R$ and $\mathrm{RSS} = 0.081(32)$ for $\tilde{Z}_T$ -- when averaged, we find an overall $\mathrm{RSS} = 0.12(5)$. This analysis yields a better agreement than the previous analysis for $\bar{n} = 0.00$, however, we found the $\mathrm{RSS}$ to overlap, within their margin of errors.

\subsubsection{Comparison of error models}
\noindent
To conclude this section we compare the individual models via their respective RSS in Tab~\ref{tab:app_model_comparison_rss}: heating of the phonon mode with a rate $\Gamma_{\mathrm{H}} = 27(1)$ phonons per second and a model considering dephasing of the qubits with $p_z = 0.03$, neglecting $X$ and $Y$ errors. We consider both cases of a perfectly cooled initial state $\bar{n} = 0.00$ and a finite temperature state $\bar{n} = 0.05$. As it is evident from both models, the initial temperature in the considered range has only a small effect. For $\bar{n} = 0.05$ we find an average $\mathrm{RSS} = 0.53(14)$ assuming errors are only related to heating of the phonon mode, compared to $\mathrm{RSS} = 0.12(5)$ for the weighted Pauli model. Dephasing of the qubits is attributed to various mechanisms: first, heating yields to the occupation of different motional states, each evolving with a different frequency depending on the number of phonons. Second, fluctuations of the tip voltages of the trap electrodes induce phases on the COM mode, in turn inducing a phase shift on the qubits upon each sideband operation. A third mechanism is imperfect compensation of the Stark shift, resulting in a small detuning from the first blue sideband of the COM mode. All of these sources manifest as $Z$ errors. As increasing the system size $N$ extends the length of the circuit while simultaneously decreasing the coherence time of the COM mode, these errors will become evermore relevant when adding more qubits. In contrast, systematic $X$ and $Y$ errors do not impact the measurement outcomes, as the VQE tunes the rotation angles to implement the desired operations. 
\begin{table}[h!]
    \centering
    \begin{tabular}{c|c|c|c|c}
        Model & $\bar{n}$ & RSS $\tilde{Z}_R$ & RSS $\tilde{Z}_T$ & RSS (avg.)\\
        \toprule
        \multirow{2}{*}{Finite temperature} & 0.00 & 2.41(76) & 3.23(46) & 2.82(61) \\
        & 0.05 & 1.11(96) & 1.92(48) & 1.52(22) \\
        \hline
        \multirow{2}{*}{Heating} & 0.00 & 0.84(21) & 0.45(11) & 0.65(16) \\
        & 0.05 & 0.71(18) & 0.35(10) & 0.53(14) \\
        \hline
        \multirow{2}{*}{Pauli $p_z = 0.03$} & 0.00 & 0.17(7) & 0.098(35) & 0.13(5)\\
        & 0.05 & 0.15(7) & 0.081(32) & 0.12(5)
    \end{tabular}
    \caption{Comparison of the RSS obtained for the `best fit` of the considered error models. We show the values obtained for $\tilde{Z}_R$ and $\tilde{Z}_T$ as well as their average for different mean phonon numbers $\bar{n}$ of the initial state.}
    \label{tab:app_model_comparison_rss}
\end{table}
\section{A 4 qubit SSH test bed}
\label{app:app_four_qubit_sequence}
\noindent
We test our blue sideband sequences on the smallest instance of the SSH model, namely on a system of four qubits. The circuit follows the scheme shown in Fig.~\ref{fig:fig1_circuit_ions}(a). However, the final `box' $\boldsymbol\theta_D$ is executed right after $\boldsymbol\theta_C$ on qubits $3$ and $4$. We obtain the optimal parameter set $\boldsymbol\theta_{\mathrm{opt.}}$ from theory; the 14 angles (in units of $\pi$) required to implement the circuit are given in Tab. \ref{tab:angles_four_qubits} below.
\begin{table}[h!]
    \centering
    \begin{tabular}{c|c|c|c|c|c}
    ~Box~ & \multicolumn{5}{c}{Angles $[\pi]$}\\
    \toprule
    $\boldsymbol\theta_A$ & 1.2036 & -0.3984 & & &\\
    $\boldsymbol\theta_B$ & 0.1526 & 0.9366 & -1.1738 &  0.067 & -0.0562\\
    $\boldsymbol\theta_C$ & -0.9232 & 1.5904 & -0.1288 & -1.0344 & -2.8254\\
    $\boldsymbol\theta_D$ & 0.1792 & 3.8938 & & & 
    \end{tabular}
    \caption{Rotation angles for the individual boxes of the circuit shown in Fig.~\ref{fig:fig1_circuit_ions}(a) for the 4 qubit test bed system; all angles are given in units of $\pi$.}
    \label{tab:angles_four_qubits}
\end{table}
\section{Proposed implementation of a 2D TN-VQE in a novel ion trap}
\label{app:2Dproposal}
\begin{figure*}[t]
    \centering
    \includegraphics[scale=0.6]{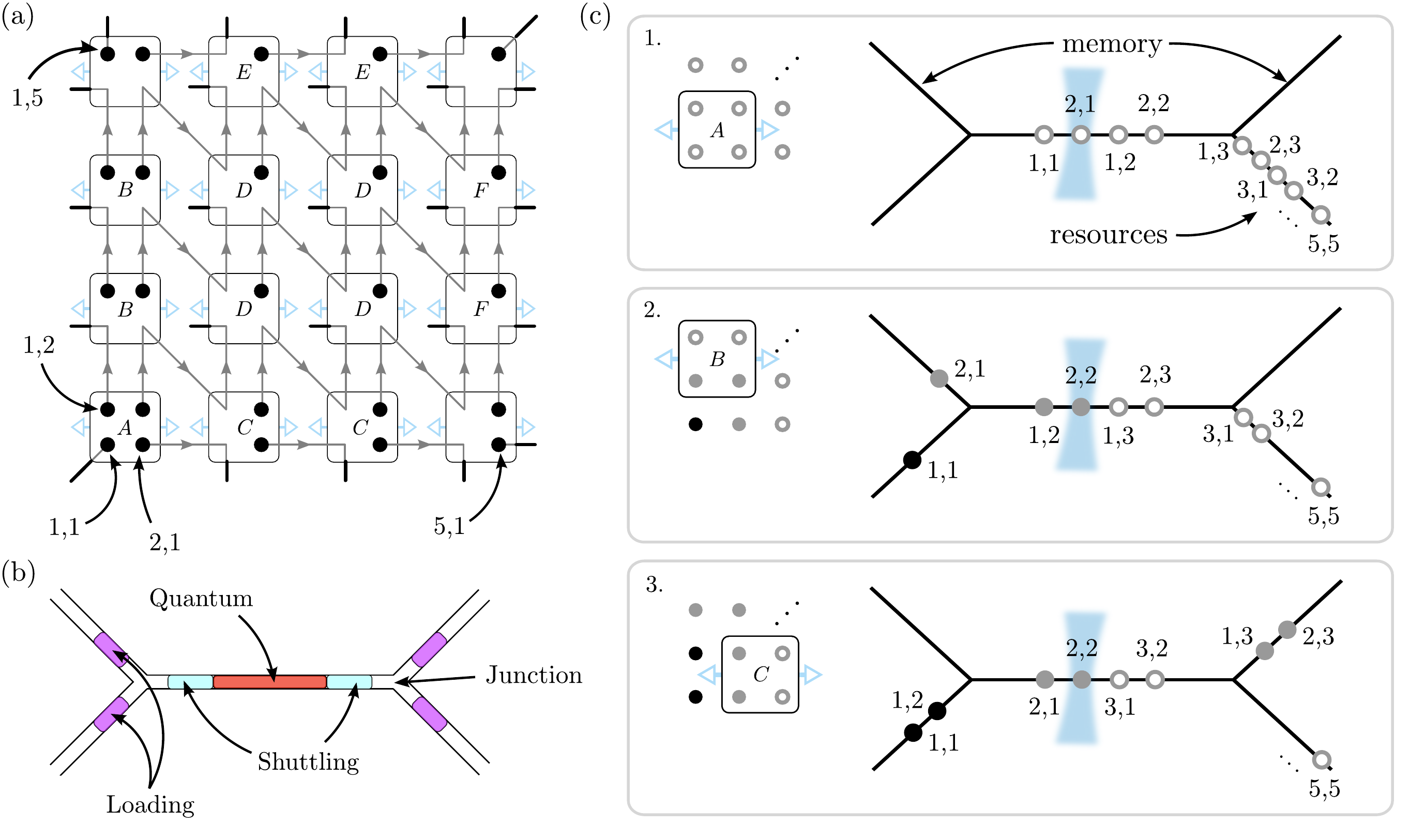}
    \caption{
    Tensor-Network-VQE in a possible architecture of a 2D ion trap.  (a) TN representation of a 2D circuit in a system of $5\times 5$ ions (circles). Each plaquette couples four qubits to a local phonon mode via sideband operations similarly to Fig.~\ref{fig:fig1_circuit_ions}. Plaquettes labeled by identical letters can share variational parameters to enforce the approximate translational invariance. 
    (b) A suitable linear trap architecture \citep{osti_1237003} -- the quantum region in the center is connected to four branches, which serve as loading and memory zones. Ions can be shuttled in between regions by precise control of local trap potentials. (c) Possible implementation of the 2D circuit via sideband and shuttling operations -- we show the first three steps in the circuit with all participating ions shuttled to their respective positions. Here, empty grey circles indicate ions that have thus far not participated in the circuit, while those in solid grey have already been addressed and must be kept for later use in different plaquettes. Black circles highlight ions that have already dropped out of the circuit and remain from this point on untouched.
    }
    \label{fig:app_2D_Network}
\end{figure*}

\noindent
In this section we speculate on a potential preparation of plaquette-PEPS~\cite{wei2021sequential} with two-dimensional (2D) TN-VQE, using a novel quantum technology of linear ion trap as quantum hardware \citep{osti_1237003}. We demonstrate a minimal scheme, involving 25 ions, which can exploit bulk-translation invariance in the target state; the TN diagram is shown in Fig. \ref{fig:app_2D_Network}(a). Similar to the MPS diagram in Fig. \ref{fig:fig1_circuit_ions}(c), the `plaquettes' encode a sequence of unitary operations on local sets of ions.

The architecture of choice is shown in Fig.~\ref{fig:app_2D_Network}(b) and has already been tested under experimental conditions \citep{osti_1237003}. It incorporates four `storage' branches, which are used to load and to store ions. Via `shuttling' zones the ions can be moved to the central `quantum region', where they reside in a common potential and thus share a (local) phonon mode. Similar to the scheme in Fig. \ref{fig:fig1_circuit_ions}(a/b) we use single-ion sideband operations to implement a variational unitary. However, we need to consider controlled reconfigurations of ion string(s) in the QR to realize the different plaquettes; these reconfigurations are implemented by modulation of individual trap electrodes. We employ three operations, namely (1) `shuttling' operations, where ions are moved in between different trap zones, (2) `splits', which divide an ion string into substrings and (3) `merges', where substrings are joined into one, larger string. 
\newline\newline\noindent
We will now show how the TN in Fig.~\ref{fig:app_2D_Network}(a) can be mapped to the chosen trap architecture. Considering the lower left corner of the TN diagram, we sketch the ion string reconfiguration operations required to realize the first three plaquettes $A$, $B$ and $C$ in detail in Fig.~\ref{fig:app_2D_Network}(c); optical single qubit sideband operations are not explicitly shown. 
The ions are represented by colored circles, where empty grey circuits are considered `fresh' resources, i.e., ions that have not yet participated in the circuit. Solid grey circles indicate ions that have already been addressed, but must be kept in one of the memory regions to be used in later plaquettes. Finally, solid black circles show those ions, that have already `dropped out' of the circuit; they are thus parked in the lower left branch of the trap. To better identify the movement of individual ions we assign tuples of integers $(n, m)$, which correspond to the column and row of the particle in the TN diagram. 
\begin{enumerate}
    \item We implement the first plaquette $A$ by shuttling four ions from the resource register to the QR region. Note, that the initial arrangement of the qubits is chosen as such, that the number of required reconfigurations required in later plaquettes are minimized. 
    \item Continuing with $B$, we can move qubit $(1,1)$ to the lower left branch -- this ion will not participate in subsequent plaquettes. Ion $(2,1)$ is split from the substring $(1,2)$ \& $(2, 2)$ and shuttled to the left memory. Two more ions $(1,3)$ and $(2,3)$ join the QR and merge with the aforementioned substring.
    \item Plaquette $C$ requires more reordering -- $(1,3)$ and $(2,3)$ are moved to the right memory, while $(1,2)$ is parked next to $(1,1)$. Ion $(2,1)$ is shuttled back to the QR and joins $(2,2)$ with two more ions $(3,1)$ and $(3,2)$ from the resource branch.  
\end{enumerate}
In total, we need to implement 15 reconfiguration operations to implement the first three plaquettes; the individual operations are listed in Tab.~\ref{tab:2D_shuttling} below. We extend our investigation to the full circuit, ending up with a total of 152 operations. 
\begin{table}[h!]
    \centering
    \begin{tabular}{c|c|c|c|c}
    ~Step~ & ~Shuttling~ & ~Split~ & ~Merge~ & ~Total~\\
    \toprule
    1 & 1 & 1 & 0 & 2\\
    2 & 3 & 4 & 1 & 7\\
    3 & 4 & 3 & 2 & 7\\
    \end{tabular}
    \caption{Required number of shuttling, split and merge operations to realize each of the first three plaquettes of the TN diagram for the given trap configuration, see Fig.~\ref{fig:app_2D_Network}.}
    \label{tab:2D_shuttling}
\end{table}
However, the feasibility of such demanding circuits remains an open question. Recently a fault-tolerant parity readout scheme has been realized, requiring more than 40 split-and-merge operations and 110 shuttlings including state preparation and readout \citep{hilder2021faulttolerant}. However, the trap geometry used in this experiment does not feature branches as memory regions; an implementation in the trap architecture described in this section might greatly simplify the circuit. 
\clearpage
% ******************************************************** %

\bibliographystyle{apsrev4-1}
\bibliography{refs}

\end{document}